\documentclass[prd,nofootinbib,showpacs,superscriptaddress, preprint]{revtex4}
\usepackage[T1]{fontenc}
\usepackage[english]{babel}
\usepackage[utf8x]{inputenc}
\usepackage{amsmath,amssymb}
\usepackage{epsfig}
\usepackage{graphicx}
\usepackage[usenames,dvipsnames]{color}
\usepackage{slashed}
\usepackage[colorlinks,citecolor=blue]{hyperref}
\usepackage{color}
\usepackage{subfig}
\usepackage{float}
\begin{document}
\title{Right-handed Neutrino Dark Matter with Radiative Neutrino Mass in Gauged $B-L$ Model}

\author{Debasish Borah}
\email{dborah@iitg.ac.in}
\affiliation{Department of Physics, Indian Institute of Technology
Guwahati, Assam 781039, India}
\author{Dibyendu Nanda}
\email{dibyendu.nanda@iitg.ac.in}
\affiliation{Department of Physics, Indian Institute of Technology
Guwahati, Assam 781039, India}
\author{Nimmala Narendra}
\email{ph14resch01002@iith.ac.in}
\affiliation{Indian Institute of Technology Hyderabad, Kandi, Sangareddy, 502285, Telangana, India}
\author{Narendra Sahu}
\email{nsahu@iith.ac.in}
\affiliation{Indian Institute of Technology Hyderabad, Kandi, Sangareddy, 502285, Telangana, India}

\begin{abstract}
We study the possibility of right-handed neutrino dark matter (DM) in gauged $U(1)_{B-L} \times Z_2$ extension of the standard 
model augmented by an additional scalar doublet, being odd under the $Z_2$ symmetry, to give rise the scotogenic scenario of 
radiative neutrino masses. Due to lepton portal interactions, the right-handed neutrino DM can have additional co-annihilation 
channels apart from the usual annihilations through $Z_{B-L}$ which give rise to much more allowed mass of DM from relic abundance 
criteria, even away from the resonance region like $M_{\rm DM} \approx M_{Z_{B-L}}/2$. This enlarged parameter space 
is found to be consistent with neutrino mass constraints while being sensitive to direct detection experiments of DM as well as 
rare decay experiments looking for charged lepton flavour violating decays like $\mu \rightarrow e \gamma$. Due to the possibility 
of the $Z_2$ odd scalar doublet being the next to lightest stable particle that can be sufficiently produced in colliders by virtue 
of its gauge interactions, one can have interesting signatures like displaced vertex or disappearing charged tracks provided that the mass 
splitting $\delta M$ between DM and the next to lightest stable particle (NLSP) is small. In particular, if $\delta M < m_\tau=1.77$ GeV, 
then we get large displaced vertex signature of NLSP while being consistent with neutrino mass, lepton flavour violation and observed 
relic density.
% We provide 
%the model parameters from the requirement of satisfying all the relevant bounds while having the tantalising possibility 
%of giving these interesting collider signatures.}
\end{abstract}

%\keywords{}
%%%%%%%%%%%%%%%%%%%%
\maketitle
%\flushbottom
\section{Introduction} \label{sec1}
It is quite well known, thanks to several evidences gathered in the last few decades, starting from the galaxy cluster 
observations by Fritz Zwicky \cite{Zwicky:1933gu} back in 1933, observations of galaxy rotation curves in 1970's \cite{Rubin:1970zza} 
and the more recent observation of the bullet cluster \cite{Clowe:2006eq} to the latest cosmology data provided by the Planck 
satellite \cite{Aghanim:2018eyx}, that the present Universe is composed of a mysterious, non-luminous and non-baryonic form of matter, 
known as dark matter (DM). The latest data from the Planck mission suggest that the DM constitutes around $27\%$ of the total energy 
density of the present Universe. In terms of density parameter $\Omega_{\rm DM}$ and $h = \text{(Hubble Parameter)}/(100 \;\text{km} \text{s}^{-1} 
\text{Mpc}^{-1})$, the present DM abundance is conventionally reported as \cite{Aghanim:2018eyx}:
\begin{equation}
\Omega_{\text{DM}} h^2 = 0.120\pm 0.001
\label{dm_relic}
\end{equation}
at 68\% CL. However, in spite of such overwhelming evidences from astrophysics and cosmology based experiments, very little is known about the 
particle nature of DM. The typical list of criteria, that a particle DM candidate has to satisfy \cite{Taoso:2007qk}, already rules out all 
the standard model (SM) particles from being a DM candidate. This implies that we need physics beyond the standard model (BSM) to incorporate the 
cosmic DM abundance. The most widely studied DM scenario so far has been the weakly interacting massive particle (WIMP) paradigm. Here, 
the DM particle, having mass and interactions typically in the electroweak scale, can give rise to the correct relic abundance after thermal 
freeze-out, a remarkable coincidence often referred to as the \textit{WIMP Miracle} \cite{Kolb:1990vq}. For a recent review, one may refer 
to \cite{Arcadi:2017kky}. Such electroweak scale mass and interactions make this WIMP paradigm very appealing from direct detection point 
of view as well \cite{Liu:2017drf}.

Apart from DM, another equally appealing motivation for BSM physics is the observed neutrino mass and mixing which have been confirmed 
by several experiments for more than a decade till now \cite{Fukuda:2001nk, Ahmad:2002jz, Ahmad:2002ka, Abe:2008aa, Abe:2011sj, Abe:2011fz, 
An:2012eh, Ahn:2012nd, Adamson:2013ue, Olive:2016xmw}. Among them, the relatively recent experimental results from the T2K \cite{Abe:2011sj}, 
Double Chooz \cite{Abe:2011fz}, Daya Bay \cite{An:2012eh}, RENO \cite{Ahn:2012nd} and MINOS \cite{Adamson:2013ue} experiments have not only 
confirmed the results from earlier experiments but also discovered the non-zero reactor mixing angle $\theta_{13}$. For a recent global fit 
of neutrino oscillation data, we refer to \cite{Esteban:2018azc}. Apart from neutrino oscillation experiments, the neutrino sector is constrained 
by the data from cosmology as well. For example, the latest data from the Planck mission constrain the sum of absolute neutrino masses 
$\sum_i \lvert m_i \rvert < 0.12$ eV \cite{Aghanim:2018eyx}. Similar to the observations related to DM, these experimental observations also 
can not be addressed by the SM as neutrinos remain massless at the renormalisable level. The Higgs field, which lies at the origin of all massive 
particles in the SM, can not have any Dirac Yukawa coupling with the neutrinos due to the absence of the right-handed neutrino. Even if the right 
handed neutrinos are included, one needs the Yukawa couplings to be heavily fine tuned to around $10^{-12}$ in order to generate sub-eV neutrino 
masses from the same Higgs field of the SM. At non-renormalisable level, one can generate a tiny Majorana mass for the neutrinos from the same 
Higgs field of the SM through the dimension five Weinberg operator \cite{Weinberg:1979sa}. However, the unknown cut-off scale $\Lambda$ in such 
operators points towards the existence of new physics at some high energy scale. Several BSM proposals, known as seesaw mechanisms \cite{Minkowski:1977sc, GellMann:1980vs, Mohapatra:1979ia, Schechter:1980gr}, attempt to provide a dynamical origin of such operators by incorporating additional fields. 
Apart from the conventional type I seesaw, there exist other variants of seesaw mechanisms also, namely, type II seesaw \cite{Mohapatra:1980yp, Lazarides:1980nt, Wetterich:1981bx, Schechter:1981cv, Brahmachari:1997cq}, type III seesaw \cite{Foot:1988aq} and so on.

Although the origin of neutrino mass and DM may appear to be unrelated to each other, it is highly appealing and economical to find a common origin of both. 
Motivated by this here we study a very well motivated BSM framework based on the gauged $U(1)_{B-L}$ symmetry \cite{Mohapatra:1980qe, 
Marshak:1979fm, Masiero:1982fi, Mohapatra:1982xz, Buchmuller:1991ce}, where $B$ and $L$ correspond to baryon and lepton numbers respectively. The most 
interesting feature of this model is that the inclusion of three right-handed neutrinos, as it is done in type I seesaw mechanism of generating light 
neutrino masses, is no longer a choice but arises as a minimal possible way to make the new $U(1)_{B-L}$ gauge symmetry anomaly free. \footnote{For other 
exotic and non-minimal solutions to such anomaly cancellation conditions, please refer to \cite{Montero:2007cd, Wang:2015saa,Patra:2016ofq, Nanda:2017bmi, DeRomeri:2017oxa, Bernal:2018aon} and references therein.} The model has also been studied in the context of dark matter by several groups \cite{Rodejohann:2015lca, 
Okada:2010wd, Dasgupta:2014hha, Okada:2016tci, Klasen:2016qux,Sahu:2005fe,Kohri:2013sva,Kohri:2009yn}. DM in scale invariant versions of this model 
was also studied by several authors \cite{Okada:2012sg, Guo:2015lxa}. Although the scalar DM in such models can be naturally stable by virtue of 
its $B-L$ charge, the fermion DM can not be realised in the minimal model except for the possibility of a keV right-handed neutrino DM which is cosmologically long lived \cite{Biswas:2018iny}. One can introduce additional discrete symmetries, such as $Z_2$ that can stabilise one of the right-handed neutrinos \cite{Basak:2013cga, Okada:2016gsh, Okada:2018ktp, Escudero:2018fwn} while the other two neutrinos take part in the usual type I seesaw mechanism, giving rise to solar and atmospheric neutrino mixing. Since the right-handed neutrino DM in this case annihilates into the SM particles only through the $U(1)_{B-L}$ gauge bosons, the relic density is typically satisfied only near the resonance $M_{\rm DM}\approx M_{Z_{B-L}}/2$. Since the experimental limits from LEP II constrain such new gauge sector by giving a lower bound on the ratio of new gauge boson mass to the corresponding gauge coupling $M_{Z_{B-L}}/g_{B-L} \geq 7$ TeV \cite{Carena:2004xs, Cacciapaglia:2006pk}, typically one gets a lower bound on $Z_{B-L}$ mass to be around 3 TeV for generic gauge coupling $g_{B-L}$ similar to electroweak gauge couplings. This constrains the allowed DM mass to be more than a TeV. Presence of additional light scalars can however, allow lighter DM as well. But in this case also, the allowed DM mass should lie in the vicinity of the resonance region. Apart from this close tuning of DM mass depending upon the mediator masses, the DM sector also gets decoupled from the neutrino mass generation mechanism in such a case, due to the absence of any coupling of DM to the leptons.

In this work, we consider the SM augmented by $U(1)_{B-L} \times Z_2$ symmetry. In addition to three right-handed neutrinos: $N_{iR}$, we introduce one scalar 
doublet $\eta$ which are all odd under the discrete $Z_2$ symmetry. The gauged $B-L$ symmetry is broken by introducing a singlet scalar $\chi$ which acquires a non-zero vacuum expectation value (VEV). As a result the low energy phenomenology of this model is similar to the popular BSM framework that provides a common origin of neutrino mass and DM, known as the scotogenic scenario as proposed by Ma \cite{Ma:2006km}, where the $Z_2$ odd particles take part in radiative generation of light neutrino masses. We consider the lightest right-handed neutrino to be lightest $Z_2$ odd particle and hence the DM candidate. We note that this model was proposed by the authors of \cite{Kanemura:2011vm} with limited discussions on right handed neutrino dark matter relic. In this model, we perform a more detailed study of right handed neutrino dark matter, pointing out all possible effects that can affect its relic abundance. Due to the existence of new Yukawa interactions, we find that the parameter space giving rise to correct relic abundance is much larger than the resonance region $M_{\rm DM}\approx M_{Z_{B-L}}/2$ for usual right-handed neutrino DM in $U(1)_{B-L}$ model. This is possible due to additional annihilation and co-annihilation channels that arise due to Yukawa interactions. We also check the consistency of this enlarged DM parameter space with constraints from direct detection, lepton flavour violation (LFV) as well as neutrino mass. Since the $Z_2$ odd scalar doublet can be the 
next to lightest stable particle (NLSP) in this case, it's charged component can be sufficiently produced at the Large Hadron Collider (LHC) by virtue of its electroweak gauge interactions, provided that it is in the sub-TeV regime. Due to the possibility of small mass splitting between NLSP and DM as well as within the components of the $Z_2$ odd scalar doublet, we can have interesting signatures like displaced vertex or disappearing charged track (DCT) which the LHC is searching 
for. To make the analysis coherent with the objectives, we constrain the model parameters in such a way that they agree with all relevant experimental bounds from cosmology, neutrino and flavour physics, direct detection and at the same time have the potential to show interesting signatures at the LHC. In particular, we show that if the mass splitting between the DM and NLSP is less than $\tau$ lepton mass, then we can get displaced vertex upto 10 cm. In addition to that the parameter space also remains sensitive to ongoing and near future runs of dark matter direct detection as well as rare decay experiments looking for lepton flavour violating charged lepton decay like $\mu \rightarrow e \gamma$.

This article is organised as follows. In section \ref{sec2}, we discuss the model followed by neutrino mass in section \ref{sec3}. We briefly discuss the possibility of lepton flavour violation in section \ref{sec3a} followed by the details of dark matter 
phenomenology in section \ref{sec4}. We briefly discuss some collider signatures of the model in section \ref{collider} and finally conclude in section \ref{sec6}.

\section{The Model}\label{sec2}
%%%%%%%%%%%%%%%%%%%%%%%%%%%%%%%%%%%%%%%%%%
Gauged $U(1)_{B-L}$ extension of the SM is one of the most popular BSM frameworks in the literature. Since the $B-L$ charges of all the SM fields are already known, it is very much straightforward to write the details of such a model. However uplifting the global $U(1)_{B-L}$ of the SM to a gauged one brings in unwanted chiral anomalies. This is because the triangle anomalies for both $U(1)^3_{B-L}$ and the mixed $U(1)_{B-L}-(\text{gravity})^2$ diagrams are non-zero. These triangle anomalies for the SM fermion content turns out to be
\begin{align}
\mathcal{A}_1 \left[ U(1)^3_{B-L} \right] = \mathcal{A}^{\text{SM}}_1 \left[ U(1)^3_{B-L} \right]=-3  \nonumber \\
\mathcal{A}_2 \left[(\text{gravity})^2 \times U(1)_{B-L} \right] = \mathcal{A}^{\text{SM}}_2 \left[ (\text{gravity})^2 \times U(1)_{B-L} \right]=-3
\end{align}
These anomalies can be cancelled minimally by introducing three right-handed neutrinos: $N_{iR}$ with unit lepton number each, which is exactly 
what we need in the SM for realising neutrino masses. These right-handed neutrinos contribute $\mathcal{A}^{\text{New}}_1 \left[ U(1)^3_{B-L} 
\right] = 3, \mathcal{A}^{\text{New}}_2 \left[ (\text{gravity})^2 \times U(1)_{B-L} \right] = 3$ leading to vanishing total of triangle anomalies. 
As pointed out before, there exists alternative and non-minimal ways to cancel these anomalies as well \cite{Montero:2007cd, Wang:2015saa,Patra:2016ofq, Nanda:2017bmi,Bernal:2018aon}.

We then extend the minimal gauged $U(1)_{B-L}$ model by introducing an additional $Z_{2}$ symmetry and a scalar doublet $\eta$ so that the right-handed 
neutrinos: $N_{iR}$ and $\eta$ are odd under the unbroken $Z_2$ symmetry. The BSM particle content of the model is shown in table \ref{tab}. The $SU(2)_L$ 
singlet scalar field $\chi$ is introduced in order to break the $U(1)_{B-L}$ gauge symmetry spontaneously after acquiring a non-zero vacuum expectation value (VEV). 
Due to the imposed $Z_2$ symmetry the neutrinos can not acquire masses at tree level, making way for radiative neutrino masses as we discuss in the next section.

\begin{table}[htbp]
\caption{New particles and their quantum numbers under the imposed symmetry.}
%\begin{center}
\begin{tabular}{|c|c|c|c|c|c|}
\hline
Fields &  SU(3)$_c$ & SU(2)$_L$ & U(1)$_Y$ & U(1)$_{B-L}$ & Z$_2$  \\
\hline
$N_{R}$ & 1 & 1 & 0 & -1 & - \\
\hline
$\chi$ & 1 & 1 & 0 & 2 & + \\
\hline
$\eta$ & 1 & 2 & $\frac{1}{2}$ & 0 & - \\
\hline
\end{tabular}
%\end{center}
\label{tab}
\end{table}

The corresponding Lagrangian can be written as:
\begin{eqnarray}
\mathcal{L} &\supseteq& \sum_{j,k=1}^{3} - y_{jk} \overline{\ell}_{jL}N_{kR} \ \tilde{\eta} - \lambda_{jk} (\overline{N_{jR})^c} \ N_{kR} \ \chi + h.c-V(H,\chi,\eta)
\end{eqnarray}
Where
\begin{eqnarray}\nonumber
V(H,\chi,\eta) &=& -\mu_H^2 H^{\dagger}H + \lambda_H (H^{\dagger}H)^2 - \mu_\chi^2 \chi^{\dagger}\chi+ \lambda_{\chi} (\chi^{\dagger}\chi)^2 +\mu_\eta^2 \eta^{\dagger}\eta + \lambda_{\eta} (\eta^{\dagger}\eta)^2  \nonumber\\
&& + \lambda_{H\chi} (H^{\dagger}H)(\chi^{\dagger}\chi)+ \lambda_{H\eta} (H^{\dagger}H) (\eta^{\dagger}\eta) + \lambda_{\chi \eta} (\chi^{\dagger}\chi)(\eta^{\dagger}\eta) \nonumber\\
&& + \lambda_1 (\eta^{\dagger}H)(H^{\dagger}\eta)+\frac{\lambda_2}{2}\bigg[ (H^{\dagger}\eta)^2+h.c.\bigg]
\end{eqnarray}
We consider the mass squared term $\mu_\eta^2>0$ so that the neutral component of only $H, \chi$ acquire non-zero VEV's v and u respectively. Expanding around the VEV, we can write the fields as:
\begin{equation}
H= \begin{bmatrix}  0  \\
   \frac{v+h}{\sqrt{2}}  
   \end{bmatrix}
\,\,\,,\,\,\,
 \chi=\frac{u+s}{\sqrt{2}}
\,\,\,\,\,{\rm and}\,\,\,\,\,
\eta= \begin{bmatrix}
      \eta^+ \\
      \frac{\eta R+i \eta I}{\sqrt{2}} 
      \end{bmatrix}.
\end{equation}

The minimisation conditions of the above scalar potential will give 
\begin{eqnarray}\nonumber
\mu_{H}^{2} = \lambda_{H} v^{2} + \frac{1}{2} \lambda_{H \chi} u^2 \\
\mu_\chi^{2} = \lambda_{\chi} u^2 + \frac{1}{2} \lambda_{H \chi} v^2
\end{eqnarray}

As a result the neutral scalar mass matrix becomes:
\begin{equation}
M^{2}(h,s)=  \begin{bmatrix}
    2 \lambda_{H} v^2  &  \lambda_H \chi u v \\
    \lambda_{H \chi} u v  &  2 \lambda_{\chi}  u^2 
  \end{bmatrix}.
\end{equation}
The mass eigenstates h$_{1}$ and h$_2$ are linear combinations of
h and s and can be written as
\begin{eqnarray}
h_{1}= h \cos \gamma - s \sin \gamma \\
h_{2}= h \sin \gamma + s \cos \gamma
\end{eqnarray}
where 
\begin{eqnarray}
\tan 2 \gamma=\frac{\lambda_{H \chi} u v}{\lambda_{\chi} u^{2}-\lambda_{H} v^{2}}.
\label{Hphimixing}
\end{eqnarray}
Such a mixing can be tightly constrained by LEP as well as LHC Higgs exclusion searches as shown recently by \cite{Dupuis:2016fda}. These constraints are more strong for low mass scalar and the upper bound on the mixing angle can be as low as $ \sin{\gamma} < 0.1$ \cite{Dupuis:2016fda}. We consider a conservative upper limit on the mixing parameter $ \sin{\gamma}  \leq 0.1$ for our analysis. This can be easily satisfied by suitable tuning of the parameters involved in the expression for mixing given in \eqref{Hphimixing}.

Physical masses at tree level for all the scalars can be written as:
\begin{eqnarray}
M_{h_1}^2 &=& \lambda_{H} v^2+\lambda_{\chi} u^2 + \sqrt{(\lambda_{H} v^2-\lambda_{\chi} u^2)^2+(\lambda_{H \chi} u v)^2}\\
M_{h_2}^2 &=& \lambda_{H} v^2+\lambda_{\chi} u^2 - \sqrt{(\lambda_{H} v^2-\lambda_{\chi} u^2)^2+(\lambda_{H \chi} u v)^2}\\
M_{\eta^{\pm}}^2 &=& \mu_{\eta}^2 + \frac{1}{2}\lambda_{H \eta} v^2 +\frac{1}{2}\lambda_{\chi \eta} u^2 \hspace{3.75cm}\\
M_{\eta R}^2 &=& \mu_{\eta}^2 + \frac{1}{2}(\lambda_{H \eta}+ \lambda_{1}+\lambda_{2} ) v^2+\frac{1}{2} \lambda_{\chi \eta} u^2\\
M_{\eta I}^2 &=&\mu_{\eta}^2 + \frac{1}{2}(\lambda_{H \eta}+ \lambda_{1}-\lambda_{2} ) v^2+\frac{1}{2} \lambda_{\chi \eta} u^2.
\end{eqnarray}
Thus, the scalar sector consists of one SM Higgs like scalar $h_1$, one singlet scalar $h_2$, one charged scalar $\eta^{\pm}$, another neutral scalar $\eta_R$ and one pseudoscalar $\eta_I$.

\section{Neutrino Mass}
\label{sec3}
\begin{figure}[H]
				\centering
				\includegraphics[width = 60mm]{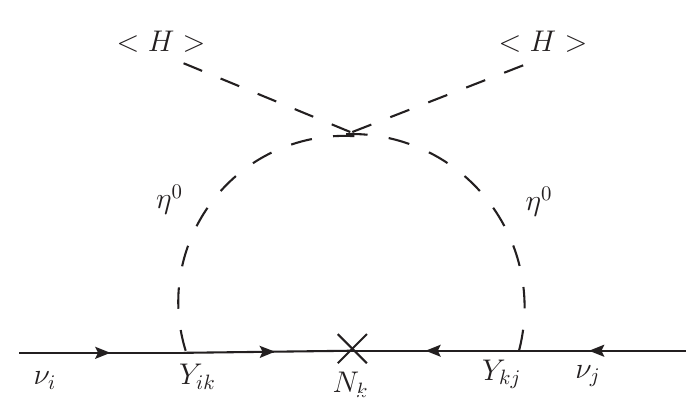}
				 \caption{\footnotesize{Radiative neutrino mass in scotogenic fashion in gauged $U(1)_{\rm B-L}$ model.}}
                \label{neutrino_mass}
\end{figure}

As mentioned earlier, neutrinos do not acquire mass through Yukawa couplings of the type $\overline{N_{R}}\tilde{H}^{\dagger} \ell$ as they are forbidden by the unbroken $Z_2$ symmetry. Therefore, type I seesaw is forbidden here. However, the term: $\frac{\lambda_2}{2}(H^{\dagger}\eta)^2$ allows us to get radiative neutrino mass at one loop level, as shown by the Feynman diagram in figure \ref{neutrino_mass},
By the exchange of Re($\eta^{0}$) and Im($\eta^{0}$) we can analytically calculate the one-loop diagram similar to \cite{Ma:2006km} which gets a non-zero contribution after the electroweak symmetry breaking $ \lambda_{2}v^{2}= M_{\eta R}^{2}-M_{\eta I}^{2}$. In our analysis we use $\lambda_2 \sim 10^{-10}$ to get the correct neutrino 
mass. 

The one-loop expression for neutrino mass is
\begin{equation}
(m_{\nu})_{ij}=\sum_{k} \frac{y_{ik} y_{kj} M_{k}}{32 \pi^{2}} \left[\frac{M_{\eta R}^{2}}{M_{\eta R}^{2}-M_{k}^{2}} \log{\left(\frac{M_{\eta R}^{2}}{M_{k}^{2}}\right)}-\frac{M_{\eta I}^{2}}{M_{\eta I}^{2}-M_{k}^{2}} \log\left(\frac{M_{\eta I}^{2}}{M_{k}^{2}}\right)\right]
\label{n_mass}
\end{equation}
where $M_{k}$ is the right handed neutrino mass.

%\subsection{Ibarra-Casas Parameterization}

The above Eq.~\eqref{n_mass} equivalently can be written as
\begin{equation}
(m_{\nu})_{ij} \equiv (y^{T}\Lambda y)_{ij}
\end{equation}
where $\Lambda$ can be defined as,
\begin{equation}
\Lambda_{k} = \frac{M_{k}}{32 \pi^{2}} \left[\frac{M_{\eta R}^{2}}{M_{\eta R}^{2}-M_{k}^{2}} \log{\left(\frac{M_{\eta R}^{2}}{M_{k}^{2}}\right)}-\frac{M_{\eta I}^{2}}{M_{\eta I}^{2}-M_{k}^{2}} \log\left(\frac{M_{\eta I}^{2}}{M_{k}^{2}}\right)\right].
\end{equation}

Since the inputs from neutrino data are only in terms of the mass squared differences and mixing angles, it is often useful to express the Yukawa couplings in terms of light neutrino parameters. This is possible through the Casas-Ibarra (CI) parametrisation \cite{Casas:2001sr} extended to radiative seesaw model \cite{Toma:2013zsa} which allows us to write the Yukawa couplings as
\begin{equation}
y=\sqrt{\Lambda}^{-1} R \sqrt{m_{\nu}} U_{\rm PMNS}^{\dagger}.
\end{equation}
Where $R$ can be a complex orthogonal matrix in general with $RR^{T}={\rm I}$. For simplicity $R$ is chosen to be real in our calculations. $U_{\rm PMNS}$ is the Pontecorvo-Maki-Nakagawa-Sakata (PMNS) leptonic mixing matrix and is given by:
\begin{equation}
U_{\text{PMNS}} = U^{\dagger}_\ell U_\nu.
\label{pmns0}
\end{equation}
If the charged lepton mass matrix is diagonal or equivalently, $U_{\ell } = {\rm I} $, then the PMNS mixing matrix is identical to the diagonalising matrix of neutrino mass matrix. 
The PMNS mixing matrix can be parametrised as
\begin{equation}
U_{\text{PMNS}}=U_\nu= \left(\begin{array}{ccc}
c_{12}c_{13}& s_{12}c_{13}& s_{13}e^{-i\delta}\\
-s_{12}c_{23}-c_{12}s_{23}s_{13}e^{i\delta}& c_{12}c_{23}-s_{12}s_{23}s_{13}e^{i\delta} & s_{23}c_{13} \\
s_{12}s_{23}-c_{12}c_{23}s_{13}e^{i\delta} & -c_{12}s_{23}-s_{12}c_{23}s_{13}e^{i\delta}& c_{23}c_{13}
\end{array}\right) U_{\text{Maj}}
\label{matrixPMNS}
\end{equation}
where $c_{ij} = \cos{\theta_{ij}}, \; s_{ij} = \sin{\theta_{ij}}$ and $\delta$ is the leptonic Dirac CP phase. 
The diagonal matrix $U_{\text{Maj}}=\text{diag}(1, e^{i\alpha}, e^{i(\beta+\delta)})$  contains the Majorana 
CP phases $\alpha, \beta$ which remain undetermined at neutrino oscillation experiments. We summarise the $3\sigma$ global fit values in table \ref{tabglobalfit} from the recent analysis \cite{Esteban:2018azc}, which we use in our subsequent analysis. Although there is some preference towards non-trivial values of Dirac CP phase in global fit data, we simply use vanishing Dirac as well as Majorana CP phases in our numerical analysis.

\begin{table}[htb]
\centering
\begin{tabular}{|c|c|c|}
\hline
Parameters & Normal Hierarchy (NH) & Inverted Hierarchy (IH) \\
\hline
$ \frac{\Delta m_{21}^2}{10^{-5} \text{eV}^2}$ & $6.79-8.01$ & $6.79-8.01 $ \\
$ \frac{|\Delta m_{31}^2|}{10^{-3} \text{eV}^2}$ & $2.427-2.625$ & $2.412-2.611 $ \\
$ \sin^2\theta_{12} $ &  $0.275-0.350 $ & $0.275-0.350 $ \\
$ \sin^2\theta_{23} $ & $0.418-0.627$ &  $0.423-0.629 $ \\
$\sin^2\theta_{13} $ & $0.02045-0.02439$ & $0.02068-0.02463 $ \\
$ \delta (^\circ) $ & $125-392$ & $196-360$ \\
\hline
\end{tabular}
\caption{Global fit $3\sigma$ values of neutrino oscillation parameters \cite{Esteban:2018azc}.}
\label{tabglobalfit}
\end{table}

\section{Lepton Flavour Violation}
\label{sec3a}
Charged lepton flavour violation arises in the SM at one loop level and remains suppressed by the smallness of neutrino masses, much beyond the current and near future experimental sensitivities. Therefore, any experimental observation of such processes is definitely a sign of BSM physics, like the one we are studying here. In the present model, this becomes inevitable due to the couplings of new $Z_2$ odd particles to the SM lepton doublets. The same fields that take part in the one-loop generation of light neutrino mass, as shown in figure \ref{neutrino_mass}, can also mediate charged lepton flavour violating processes like $\mu \rightarrow e \gamma $. The neutral scalar in the internal lines of figure \ref{neutrino_mass} will be replaced by their charged counterparts (which emit a photon) whereas the external fermion legs can be replaced by $\mu, e$ respectively, giving the one-loop contribution to $\mu \rightarrow e \gamma $. Since the couplings, masses involved in this process are the same as the ones that generate light neutrino masses and play a role in DM relic abundance, we can no longer choose them arbitrarily. The branching fraction for $\mu \rightarrow e \gamma$ that follows from this one-loop contribution can be written as~\cite{Vicente:2014wga},
\begin{equation}
{\rm Br}(\mu \rightarrow e \gamma) = \frac{3(4 \pi)^{3} \alpha_{\rm em}}{4 G_{F}^{2}}|A_{D}|^{2} {\rm Br}(\mu \rightarrow e \nu_{\mu} \bar{\nu_{e}}).
\label{br_meg}
\end{equation}
Where $\alpha_{\rm em}$ is the electromagnetic fine structure constant, $e$ is the electromagnetic coupling and $G_{F}$ is the Fermi constant. $A_{D}$ is the dipole form factor given by
\begin{equation}
A_{D} = \sum_{i=1}^{3} \frac{y_{ie}^{*}y_{i\mu}}{2(4\pi)^{2}} \frac{1}{m_{\eta^{+}}^{2}}\left(\frac{1-6 \xi_{i}+3 \xi_{i}^{2}+2 \xi_{i}^{3}-6 \xi_{i}^{2} log\xi_{i}}{6(1-\xi_{i})^{4}}\right).
\end{equation}
Here the parameter $\xi_{i}$'s are defined as $\xi_{i} \equiv M_{N_{i}}^{2}/m_{\eta^{+}}^{2}$. The MEG experiment provides the most stringent upper limit on the branching ratio ${\rm Br}(\mu \rightarrow e \gamma) < 5.7 \times 10^{-13}$~\cite{Adam:2013mnn}. A more recent bound from the same MEG collaboration that appeared in 2016 is: ${\rm Br}(\mu \rightarrow e \gamma) < 4.2 \times 10^{-13}$ \cite{TheMEG:2016wtm}.

%With different set of choices for Yukawa couplings $y_{ij}$ we try to fit the observed light neutrino masses and mixing \cite{Esteban:2016qun} as given below. A typical choice of Yukawa couplings is 
%\begin{equation}
%\begin{aligned}
%y_{e1}=0.00001, y_{e2}=0.621, y_{e3}=0.0001, y_{\mu_1}=0.00001, \\
%y_{\mu_{2}}=1.15, y_{\mu_{3}}=0.485, y_{\tau_1}=0.045, y_{\tau_{2}}=0.3, y_{\tau_{3}}=0.765.
%\end{aligned}
%\label{y values}
%\end{equation}
%The choices for mass parameters are
%\begin{align}
%M_{N_{1}}=99 \; \text{GeV}, M_{N_{2}}=10000  \; \text{GeV}, M_{N_{3}}=1000  \; \text{GeV},
%M_{\eta_{R}}=106.00000001  \; \text{GeV}, \nonumber \\
%M_{\eta_{I}}=106  \; \text{GeV}, M_{\eta R}^{2}-M_{\eta I}^{2}=2.12\times10^{-6}  \; \text{GeV}.
%\end{align}
%Thus the light neutrino mass matrix compatible with two mass squared differences and three mixing angles is given (in eV unit) as:
%\[
%m_{\nu}=
%  \begin{bmatrix}
%     0.00419 & 0.00776 & 0.00202 \\
%     0.00776 & 0.02567 & 0.02158 \\
%     0.00202 & 0.02158 & 0.02923
%  \end{bmatrix}\,.
%\]
%Note that in the above calculation, the CP phases are taken to be zero.

\section{Dark Matter}\label{sec4}
%%%%%%%%%%%%%%%%%%%%%%%%%%%%%%%%%%%%%%%%%%
The relic abundance of a dark matter ($\rm DM$) particle, which was in thermal equilibrium in the early Universe, can be calculated by solving 
the required Boltzmann equation:
\begin{equation}
\frac{dn_{\rm DM}}{dt}+3Hn_{\rm DM} = -\langle \sigma v \rangle (n^2_{\rm DM} -(n^{\rm eq}_{\rm DM})^2)
\end{equation}
where $n_{\rm DM}$ is the number density of $\rm DM$, $n^{\rm eq}_{\rm DM}$ is the equilibrium number density of $\rm DM$, $H$ is the Hubble expansion rate of the Universe and $ \langle \sigma v \rangle $ is the thermally averaged annihilation cross section of $\rm DM$. In terms of partial wave expansion 
one can write, $ \langle \sigma v \rangle = a +b v^2$. Numerical solution of the above Boltzmann equation gives \cite{Kolb:1990vq,Scherrer:1985zt}
\begin{equation}
\Omega_{\rm DM} h^2 \approx \frac{1.04 \times 10^9 x_F}{M_{\text{Pl}} \sqrt{g_*} (a+3b/x_F)}
\end{equation}
where $x_F = M_{\rm DM}/T_F$, $T_F$ is the freeze-out temperature, $M_{\rm DM}$ is the mass of dark matter, $g_*$ is the number of relativistic degrees of 
freedom at the time of freeze-out and and $M_{\text{Pl}} \approx 2.4\times 10^{18}$ GeV is the Planck mass. Dark matter particles with electroweak scale mass and couplings freeze out at temperatures approximately in the range $x_F \approx 20-30$. More generally, $x_F$ can be calculated from the relation 
\begin{equation}
x_F = \ln \frac{0.038gM_{\text{Pl}}M_{\rm DM}<\sigma v>}{g_*^{1/2}x_F^{1/2}}
\label{xf}
\end{equation}
which can be derived from the equality condition of DM interaction rate $\Gamma = n_{\rm DM} \langle \sigma v \rangle$ with the rate of expansion of the Universe $H \approx g^{1/2}_*\frac{T^2}{M_{Pl}}$. There also exists a simpler analytical formula (for s-wave annihilation) for the approximate DM relic abundance \cite{Jungman:1995df}
\begin{equation}
\Omega_{\rm DM} h^2 \approx \frac{3 \times 10^{-27} cm^3 s^{-1}}{\langle \sigma v \rangle}
\label{eq:relic}
\end{equation}
The thermal averaged annihilation cross section $\langle \sigma v \rangle$ is given by \cite{Gondolo:1990dk}
\begin{equation}
\langle \sigma v \rangle = \frac{1}{8 M_{\rm DM}^4 T K^2_2(M_{\rm DM}/T)} \int^{\infty}_{4 M_{\rm DM}^2}\sigma (s-4 M_{\rm DM}^2)\surd{s}K_1(\surd{s}/T) ds
\end{equation}
where $K_i$'s are modified Bessel functions of order $i$ and $T$ is the temperature.

If there exists some additional particles having mass difference close to that of DM, then they can be thermally accessible during the epoch of DM freeze out. This can give rise to additional channels through which DM can co-annihilate with such additional particles and produce SM particles in the final states. This type of co-annihilation effects on dark matter relic abundance were studied by several authors 
in \cite{Griest:1990kh, Edsjo:1997bg,Bell:2013wua,Bhattacharya:2015qpa,Chatterjee:2014vua}. Here we summarize the analysis of \cite{Griest:1990kh} for the 
calculation of the effective annihilation cross section in such a case. The effective cross section can given as 
\begin{align}
\sigma_{eff} &= \sum_{i,j}^{N}\langle \sigma_{ij} v\rangle r_ir_j \nonumber \\
&= \sum_{i,j}^{N}\langle \sigma_{ij}v\rangle \frac{g_ig_j}{g^2_{eff}}(1+\Delta_i)^{3/2}(1+\Delta_j)^{3/2}e^{\big(-x_F(\Delta_i + \Delta_j)\big)} \nonumber \\
\end{align}
where $x_F = \frac{m_{DM}}{T_F}$ and $\Delta_i = \frac{m_i-M_{\text{DM}}}{M_{\text{DM}}}$  and 
\begin{align}
g_{eff} &= \sum_{i=1}^{N}g_i(1+\Delta_i)^{3/2}e^{-x_F\Delta_i}
\end{align}
The masses of the heavier components of the inert Higgs doublet are denoted by $m_{i}$. The thermally averaged cross section can be written as
\begin{align}
\langle \sigma_{ij} v \rangle &= \frac{x_F}{8m^2_im^2_jM_{\text{DM}}K_2((m_i/M_{\text{DM}})x_F)K_2((m_j/M_{\text{DM}})x_F)} \times \nonumber \\
& \int^{\infty}_{(m_i+m_j)^2}ds \sigma_{ij}(s-2(m_i^2+m_j^2)) \sqrt{s}K_1(\sqrt{s}x_F/M_{\text{DM}}) \nonumber \\
\label{eq:thcs}
\end{align}
We first implement our model in \texttt{micrOMEGAs} package \cite{Belanger:2013oya} to calculate the relic abundance of DM, the results of which we discuss in the following subsections.

\subsection{Relic Density of $N_{1}$ DM in Minimal $U(1)_{B-L}$ Model}
\begin{figure}
				\centering
			\subfloat[]{{\includegraphics[width = 80mm]{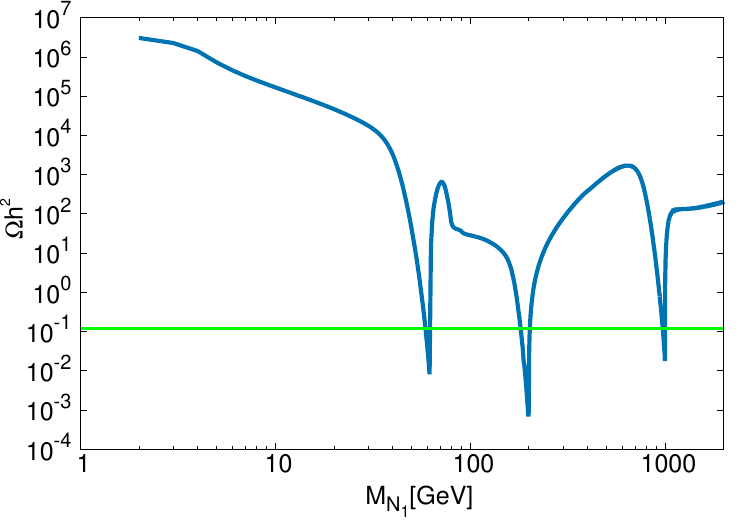} }}
%				 \caption{\footnotesize{}}
%\end{figure}
%\begin{figure}[H]
				\centering
			\subfloat[]{{\includegraphics[width = 80mm]{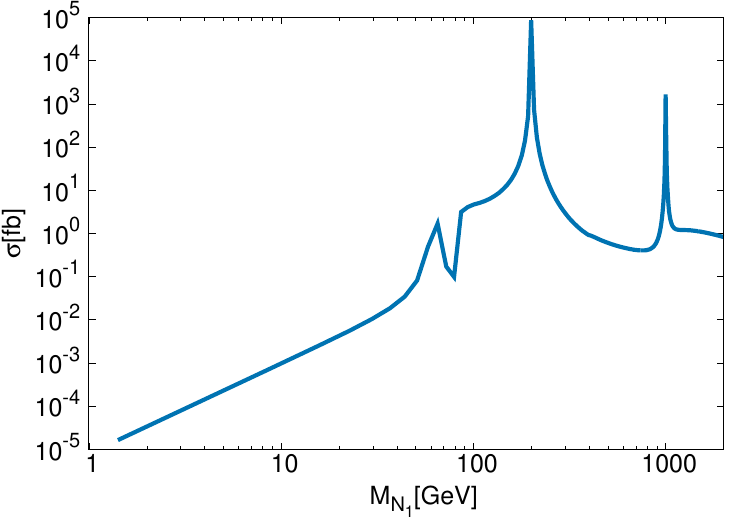} }}
				 \caption{(a) Relic density as a function of DM mass in a minimal $U(1)_{B-L}$ model. The horizontal band corresponds to the central value of Planck 2018 limit as given in Eq.\eqref{dm_relic}. (b) The annihilation cross-section of DM as a function of its mass.}
				 \label{simpleU1B-L}
\end{figure}

%%%%%%%%%%%%%%%%%%%%%%%%%%%%%%%%%%%%%%%%%%%%%%%%%%%%%%%%%%%%%%%%%%%%%%%%%
First, we show the the relic abundance of the lightest right-handed neutrino DM $N_1$ in the minimal $U(1)_{B-L}$ model so that we can later compare it with the modifications obtained in the scotogenic extension. In the minimal model, DM annihilates into SM particles either through the gauge boson $Z_{B-L}$ or 
through singlet scalar mixing with the SM Higgs as shown in figure \ref{anni} (a), (b). In figure \ref{simpleU1B-L}, we show the relic density (left) and corresponding annihilation cross-section (right) as a function of DM mass. The singlet scalar and $Z_{B-L}$ masses are taken as 400 GeV and 2 TeV respectively. The singlet-SM Higgs mixing is taken to be $\sin\gamma=0.1$ and the gauge coupling is $g_{B-L}=0.035$, in agreement with collider bounds. The three resonances corresponding to the SM Higgs, singlet scalar and the $Z_{B-L}$ boson are clearly seen in this figure. It is also clear that the correct relic abundance (corresponding to the Planck 2018 bound shown as the horizontal band in the left panel of figure \ref{simpleU1B-L}) is satisfied only near these resonance regions. This is a typical feature of fermion singlet DM in minimal $U(1)_{B-L}$ model, which we mentioned earlier.

\subsection{Relic Density of $N_{1}$ DM in Scotogenic $B-L$ model}
\begin{figure}
%    \centering
    \subfloat[]{{\includegraphics[width=60mm]{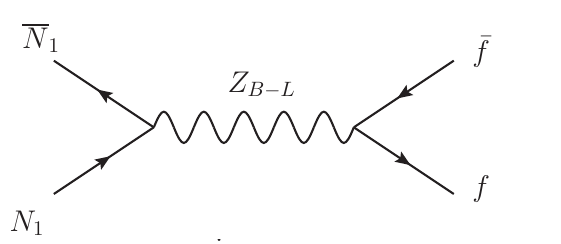} }}%
 %   \qquad
    \subfloat[]{{\includegraphics[width=60mm]{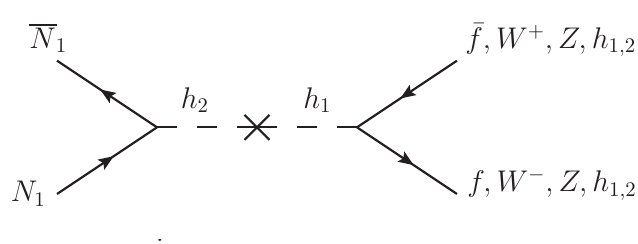} }} \\
%    \caption{}
%\end{figure}
%\begin{figure}
%   \centering
    \subfloat[]{{\includegraphics[width=50mm]{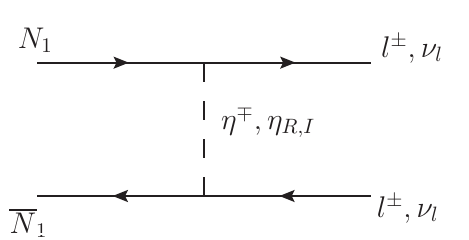} }}
    \subfloat[]{{\includegraphics[width=50mm]{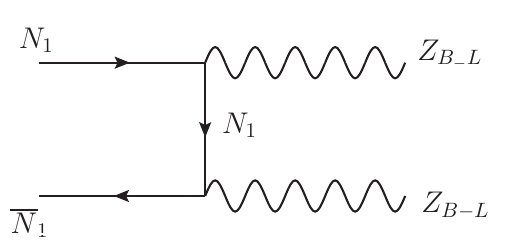} }}%
%    \qquad
    \subfloat[]{{\includegraphics[width=50mm]{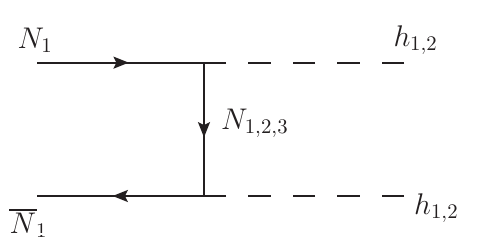} }}%
   \caption{\footnotesize{DM annihilation channels in scotogenic $U(1)_{B-L}$ model.}}
\label{anni}
\end{figure}

\begin{figure}
%    \centering
    \subfloat[]{{\includegraphics[width=60mm]{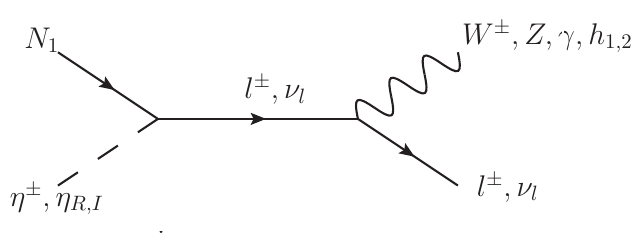} }}%
    \subfloat[]{{\includegraphics[width=60mm]{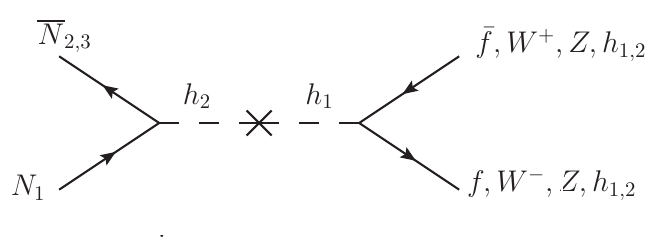} }}%
    \subfloat[]{{\includegraphics[width=50mm]{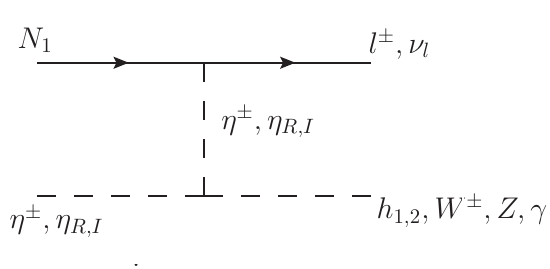} }} \\
%\end{figure}
%\begin{figure}
%    \centering
        \subfloat[]{{\includegraphics[width=50mm]{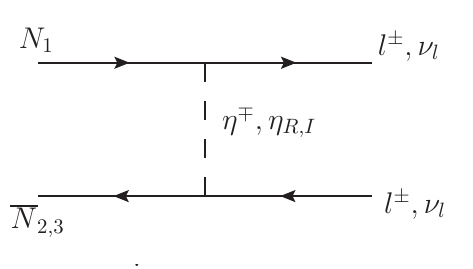} }}
      \subfloat[]{{\includegraphics[width=50mm]{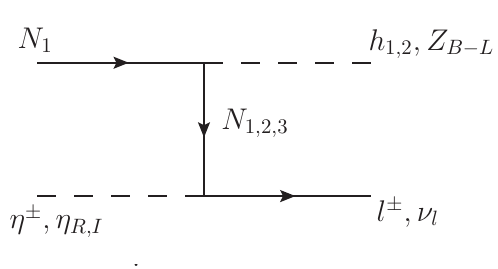} }}%
     \subfloat[]{{\includegraphics[width=50mm]{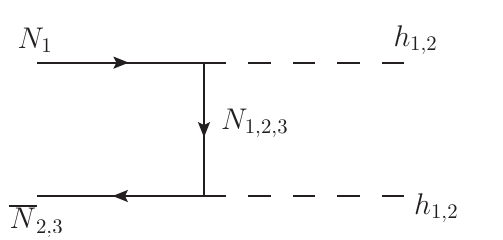} }}%
   \caption{\footnotesize{DM co-annihilation channels in scotogenic $U(1)_{B-L}$ model.}}
\label{coanni}
\end{figure}

%%%%%%%%%%%%%%%%%%%%%%%%%%%%%%%%%%%%%%%%%%%%%%%%%%%%%%%%%%%%%%%%%%%%
Apart from the usual annihilation channels of DM in minimal $U(1)_{B-L}$ model discussed above, there arises a few more annihilation and co-annihilation 
channels after extending the model in scotogenic fashion. The corresponding annihilation and co-annihilation channels are shown in figures \ref{anni} and 
\ref{coanni} respectively.

%%%%%%%%%%%%%%%%%%%%%%%%%%%%%%%%%%%%%%%%%%%%%%%%%%%%

\begin{figure}
%\begin{center}
\includegraphics[width = 70mm]{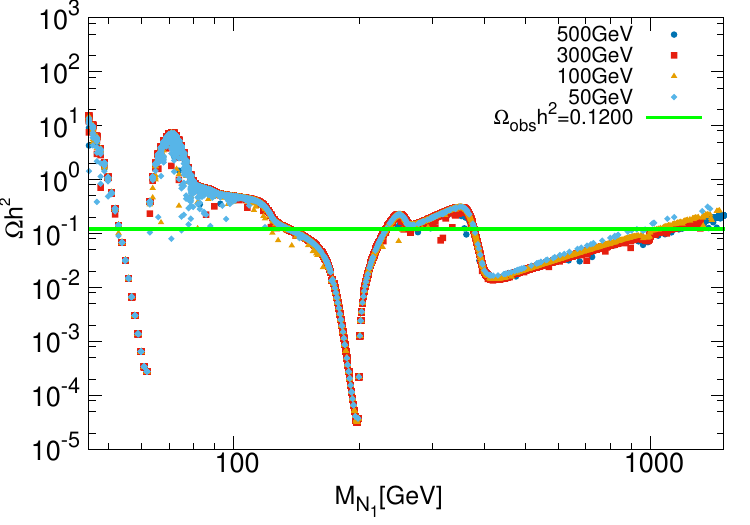}
\includegraphics[width = 70mm]{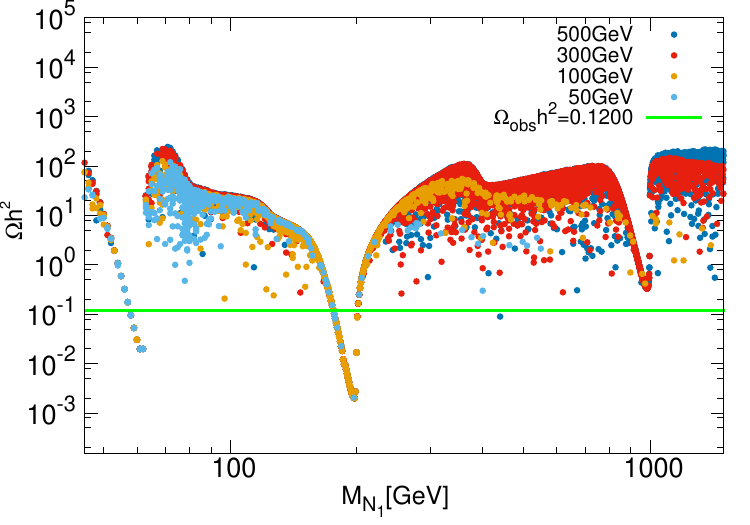}
%\end{center}
\caption{Relic density as function of DM mass for different mass splitting ($\delta M_1=M_{\eta^{\pm}, \eta_I}-M_{N_1}$) is shown in scotogenic $U(1)_{B-L}$ model. The different values of $\delta M_1= 50, 100, 300, 500$ GeV are shown respectively from sky-blue to blue points. The value of $\lambda_{11}$ is taken to be 0.9 in the left panel and 0.1 in the right panel. In all cases, we have fixed $M_{N_{2}}=M_{N_{1}}+\delta M_{2}$, $M_{N_{3}}=M_{N_{1}}+\delta M_{3}$, where $\delta M_{2}, \delta M_{3}$ are fixed at 2000 GeV and 3000 GeV respectively. The horizontal band corresponds to the central value of Planck 2018 limit as given in Eq.\eqref{dm_relic}. We use $\sin \gamma$ = 0.1, $g_{B-L}=0.035$, $M_{h_{2}}=400$ GeV and $M_{Z_{B-L}}=2$ TeV.}
\label{omg_mn1}
\end{figure}

\begin{figure}
%\begin{center}
%\includegraphics[width = 70mm]{diff_ms_mn1_omg.pdf}
%\includegraphics[width = 70mm]{diff_ms_mn1_01_omg.pdf}
\includegraphics[width = 75mm]{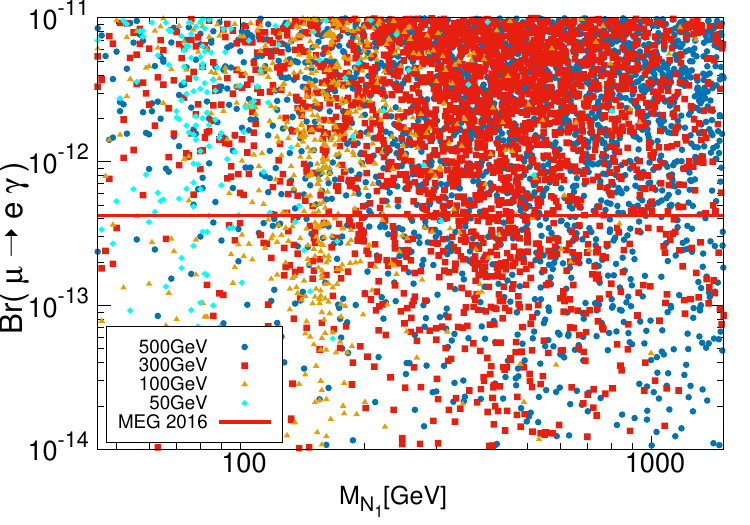}
\includegraphics[width = 75mm]{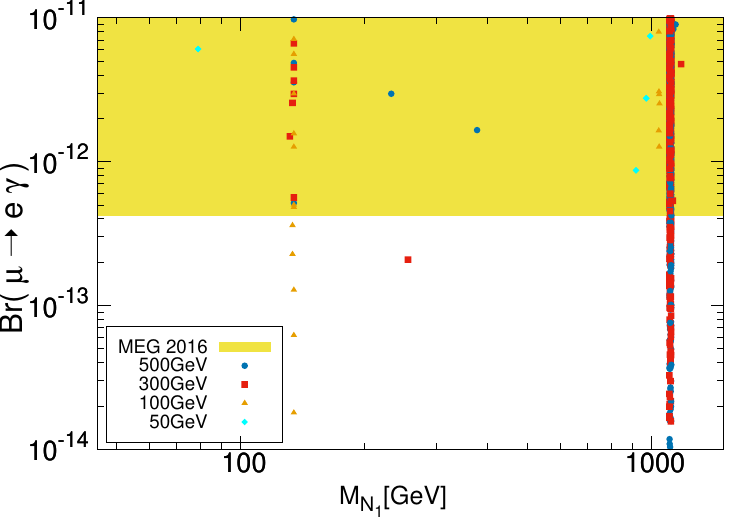}
%\end{center}
\caption{The branching fraction (left panel) as a function of DM mass for different mass splittings ($\delta M_1=M_{\eta^{\pm}, \eta_I}-M_{N_1}$) is shown in scotogenic $U(1)_{B-L}$ model. The different values of $\delta M_1= 50, 100, 300, 500$ GeV are shown respectively from sky-blue to blue points. We use $\lambda_{11}=0.9$, $\sin \gamma$ = 0.1, $g_{B-L}=0.035$, $M_{h_{2}}=400$ GeV and $M_{Z_{B-L}}=2$ TeV. In all cases, we have fixed $M_{N_{2}}=M_{N_{1}}+\delta M_{2}$, $M_{N_{3}}=M_{N_{1}}+\delta M_{3}$, where $\delta M_{2}, \delta M_{3}$ are fixed at 2000 GeV and 3000 GeV respectively. The horizontal band corresponds to the MEG 2016 upper bound ${\rm Br}(\mu \rightarrow e \gamma)$=4.2 $\times$ $10^{-13}$. In the right panel, points satisfying relic density and MEG 2016 upper bound on ${\rm Br}(\mu \rightarrow e \gamma)$ are shown for the same parameters as given above. The Yellow shaded region is not allowed by the MEG 2016 upper bound.}
\label{relic_diff_ms}
\end{figure}

%%%%%%%%%%%%%%%%%%%%%%%%%%%%%%%%%%%%%%%%%%%%%%%%%%%%%%%%%%%%%%%%%%%%%%%%%%%%%%%%%%%%%%%%%%%%%%%%%%%
We first show the effects of co-annihilations on DM relic abundance by considering four different mass splittings $\delta M_1 = M_{\rm NLSP}-M_{N_1}$ where NLSP is the scalar doublet $\eta$ and its components. In figure \ref{omg_mn1}, we show the relic abundance as a function of DM mass for $\delta M_1 = 50, 100, 300, 500$ GeV and with the singlet scalar-SM Higgs mixing $\sin\gamma=0.1$. 
The Yukawa couplings are generated through the Casas-Ibarra parametrisation so that they automatically satisfy the current experimental constraints from solar and atmospheric mass squared differences as well as mixing angles. As can be seen from figure \ref{omg_mn1}, the co-annihilation effects can change the relic abundance depending upon the mass splitting $\delta M_1$ as well as $\lambda_{11}$. We set $\lambda_{11}= 0.9$ (left panel) and $0.1$ (right panel) for the comparison purpose. We also check that these values of $\lambda_{11}$ satisfy the direct detection bounds which we will discuss in the subsequent sections. In the left panel of figure \ref{omg_mn1}, the co-annihilation effects are sub-dominant due to enhanced annihilation via singlet Higgs (caused by large coupling $\lambda_{11}$) while in the top right panel, the co-annihilation effects are visible, allowing DM mass away from the resonance regions. To generate this plot, the $h_{2}$ scalar mass and the $M_{Z_{B-L}}$ mass have been fixed at $M_{h_{2}}=400$ GeV and $M_{Z_{B-L}}=2000$ GeV respectively. The gauge coupling is fixed at $g_{B-L}=0.035$. Since the same Yukawa couplings also contribute to the charged lepton flavour violation, we compute the corresponding contribution to $\mu \rightarrow e \gamma$ using Eq.\eqref{br_meg}. The corresponding scattered plot for ${\rm Br}(\mu \to e\gamma)$ as a function of $M_{N_1}$ is shown in left panel of figure \ref{relic_diff_ms}, where the points satisfy the MEG 2016 bound on ${\rm Br}(\mu \rightarrow e \gamma)$=4.2 $\times$ $10^{-13}$. On the other hand, in the right panel of figure~\ref{relic_diff_ms}, the points satisfy the constraint from relic density as well as the MEG 2016 bound on ${\rm Br}(\mu \rightarrow e \gamma)$.

\begin{figure}
\begin{center}
\includegraphics[width = 80mm]{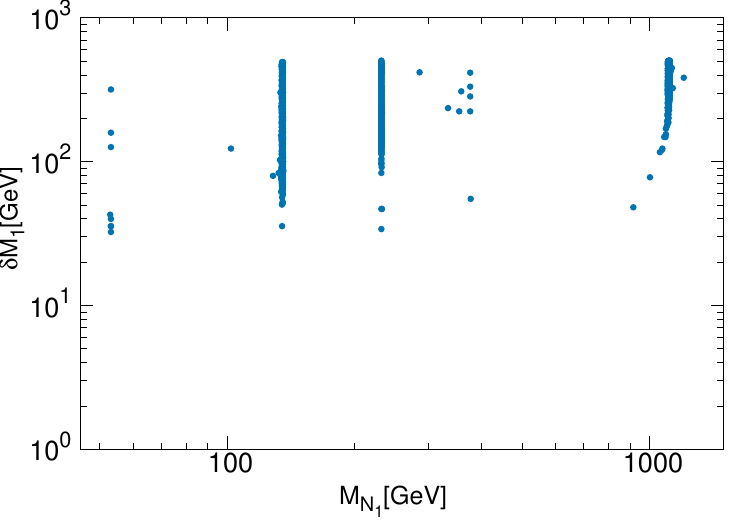}
				\end{center}
				 \caption{$\delta M_1=M_{\eta^{\pm}, \eta_I}-M_{N_1}$ versus DM mass plot at $\lambda_{11}=0.9$ which satisfy observed DM abundance and LFV constraints from $\mu \to e \gamma$. Later in figure \ref{dd2} we show that $\lambda_{11} \lesssim 0.9 $ also satisfy the stringent bound from Xenon-1T.} 
				                 \label{diff_mn1_deltm_omg}
\end{figure}

We then show the allowed parameter space in the plane of $\delta M_1$ versus $M_{N_1}$ in figure \ref{diff_mn1_deltm_omg}, 
using $\lambda_{11}=0.9$,  that satisfies the constraints from observed DM abundance, latest direct detection bound from Xenon-1T, neutrino mass as well as LFV constraints from $\mu \to e\gamma$. We see that for a given $M_{N_1}$, relic density and LFV constraints can be satisfied in a large range of $\delta M_1$. As we can see from this figure, there exists some region of parameter space around $M_{N_1} \approx M_{h_2}/2$ where almost any value of mass splitting $\delta M_1$ can satisfy the requirements due to the enhanced singlet scalar mediated resonant annihilation of DM. Similar resonance due to $Z_{B-L}$ mediation is also visible, though less prominent, near $M_{N_1} \approx M_{Z_{B-L}}/2$.

\begin{figure}
\begin{center}
				\includegraphics[width = 80mm]{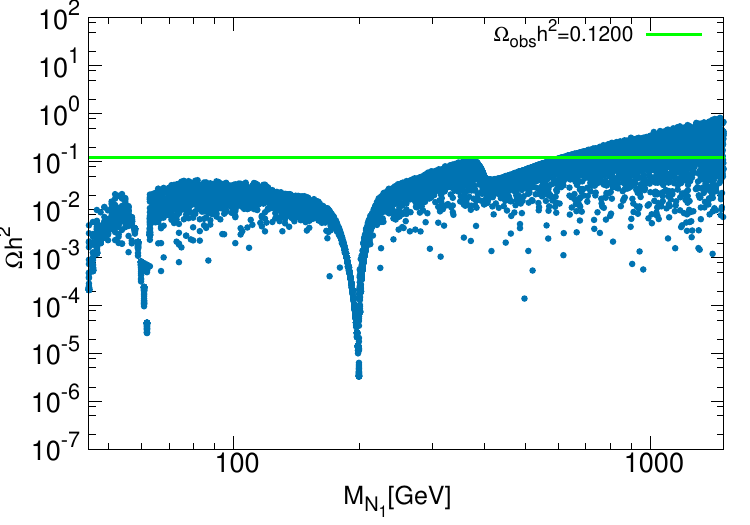}
				\includegraphics[width = 80mm]{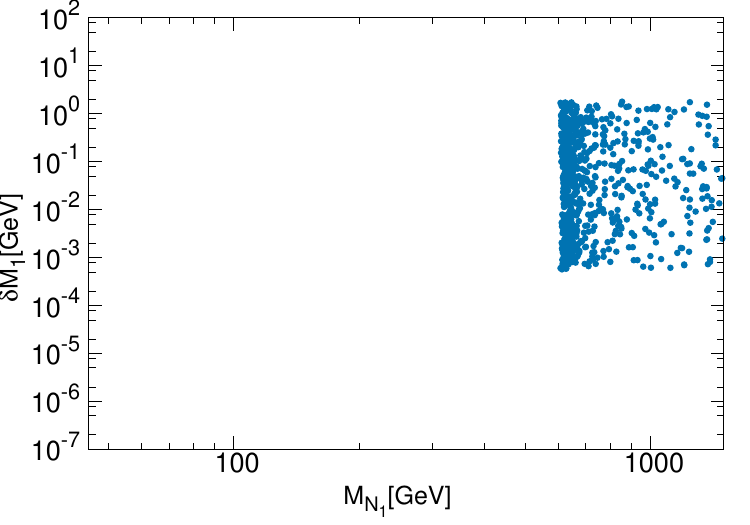}
				\end{center}
				 \caption{Left-panel: Relic abundance of DM versus its mass. Right panel: Allowed parameter space in the plane of 
$\delta M_1$ versus $M_{N_1}$ satisfying observed relic abundance from Planck and LFV from $\mu \to e \gamma$. The chosen parameters are sin$\gamma=0.1$, $M_{N_{2}}=M_{N_{1}}+ 2000$ GeV, $M_{N_{3}}=M_{N_{1}}+3000$ GeV, $M_{h_{2}}$=400 GeV, $M_{Z_{B-L}}$=2000 GeV, $g_{B-L}=0.035$, $\lambda_{11}$ = 0.9.}
               \label{mn1_omg_deltm}
\end{figure}

\begin{figure}
\begin{center}
		       \includegraphics[width = 70mm]{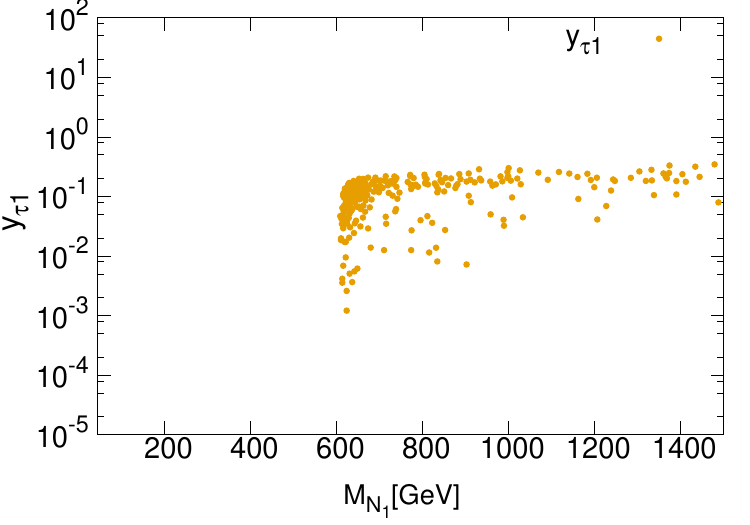}
		       \includegraphics[width = 70mm]{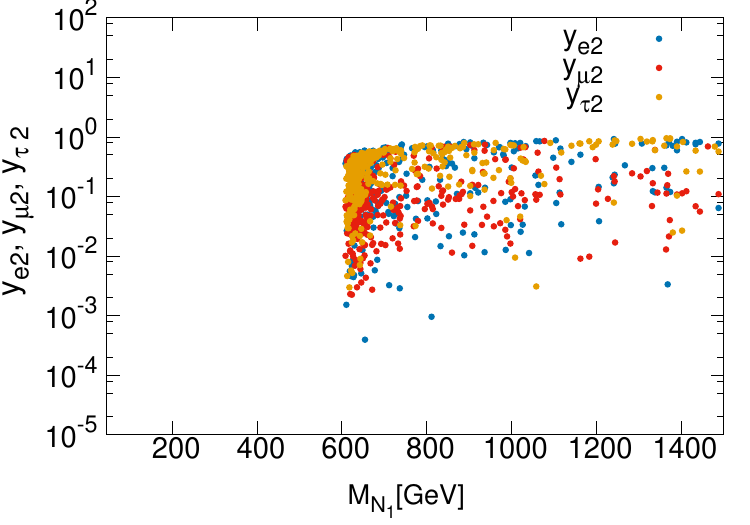}
		       \includegraphics[width = 70mm]{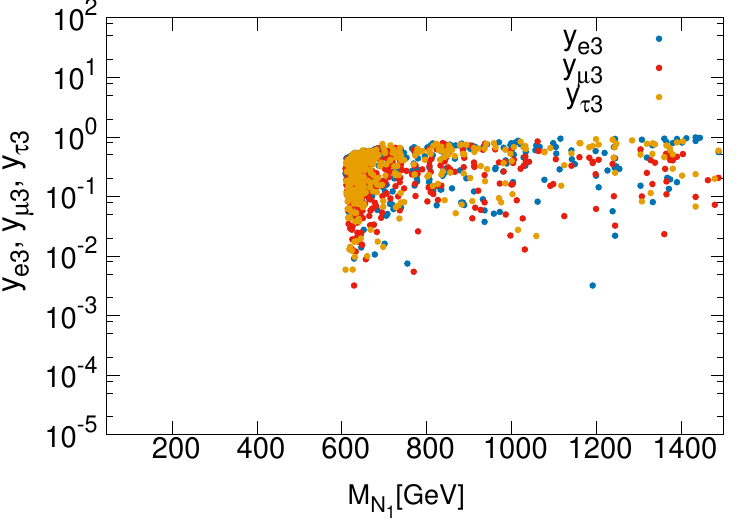}
		       \end{center}
				  \caption{The Yukawa couplings which satisfy neutrino mass, DM relic abundance and LFV constraint from $\mu \to e\gamma$ 
corresponding to the right panel of figure \ref{mn1_omg_deltm} for three generations of leptons.}
	               \label{mn1_yuk}
\end{figure}

We then consider the mass splitting $\delta M_{1}$ in the range 0.5 MeV (electron mass) to 1.77 GeV (tau mass) with 
$\delta {M_{2}}= M_{N_{2}}- M_{N_{1}} = 2000$ GeV and $\delta {M_{3}} = M_{N_{3}} - M_{N_{1}} = 3000$ GeV. Such mass splittings are chosen in light of the collider analysis that we discuss later, where we consider $\eta$-DM mass splitting ($\delta M_1$) to be less than the tau lepton mass so that $\eta^{\pm}$ can decay to first two generation leptons giving displaced vertex signatures if the Yukawa couplings are small (see section \ref{collider} for a detailed discussion). The 
corresponding results are shown in 
figure \ref{mn1_omg_deltm} with the left panel giving the relic abundance versus DM mass, while the right panel shows the parameter space in 
$\delta M_1-M_{N_1}$ plane that satisfies observed DM abundance from Planck and LFV bounds from $\mu \to e\gamma$. Note that in figure \ref{mn1_omg_deltm}, 
the scalar mixing is kept at $\sin \gamma =0.1$ and for simplicity we assume $y_{e1}=0\,,\,\, y_{\mu 1}=0$. In principle, the first two generation Yukawas 
are non-vanishing but we choose them to be small for our collider analysis (to be discussed section \ref{collider}), which in a way also helps in satisfying the lepton 
flavour violation constraint coming from {\rm Br}($\mu \rightarrow e \gamma $)=$4.2 \times 10^{-13}$.  Since such small first two generation Yukawas are anyway not going to play any significant role in DM co-annihilation we turn them off for this plot. However, the other Yukawa couplings are generated using Casas-Ibarra parametrisation so that all the points satisfy neutrino oscillation data. The Yukawa couplings which satisfy both LFV and DM relic abundance are shown in figure \ref{mn1_yuk}.

\begin{figure}
\begin{center}
		       \includegraphics[width = 80mm]{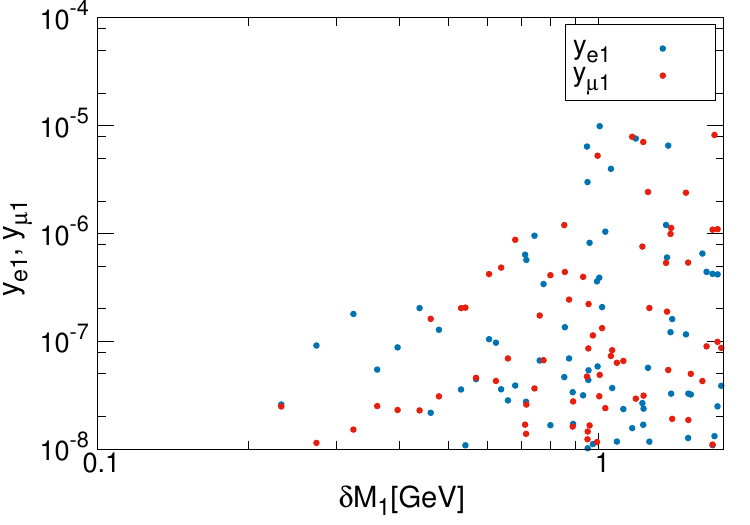}
		       \end{center}
				  \caption{Non zero values of $y_{e1}$ and $y_{\mu 1}$ which satisfy relic abundance, neutrino mass and LFV constraints.}
	               \label{deltm_yuk}
\end{figure}

The Yukawa couplings: $y_{e1}=0$ and $y_{\mu 1}=0$ are not desirable as we are looking for large displaced vertex signature of $\eta^\pm$ through the 
decay mode $\eta^\pm \to N_1 ^\pm/\mu^\pm$ (see section \ref{collider} for details). Therefore, we allow  $y_{e1}$ and $y_{\mu 1}$ to vary within the range $10^{-8}$ -$10^{-5}$, while 
other Yukawa couplings are generated through Casas-Ibarra parametrisation to obtain correct relic abundance while satisfying LFV constraints. The 
results are shown in figure \ref{deltm_yuk} in terms of $y_{e1}, y_{\mu_{1}}$ versus $\delta M_1$. We see that as $\delta M_1$ decreases we need 
smaller and smaller $y_{e1}$ and $y_{\mu 1}$ values to satisfy relic density and LFV constraints. We checked that the neutrino mass is also not 
affected for $y_{e1}, y_{\mu_{1}}\lesssim 10^{-5}$. On the other hand, such small Yukawa couplings can give rise large displaced vertex signature 
of $\eta^\pm$ as we discuss in section \ref{collider}.

%%%%%%%%%%%%%%%%%%%%%%%%%%%%%%%%%%%%%%%%%%%%%%%%%%%%%%%%%%%%%%%%%%%%%%%%%%%%%%%%%%%%%%
\subsection{Direct Detection of Dark Matter}
\label{DD}
\begin{figure}
\begin{center}
				\includegraphics[width = 40mm]{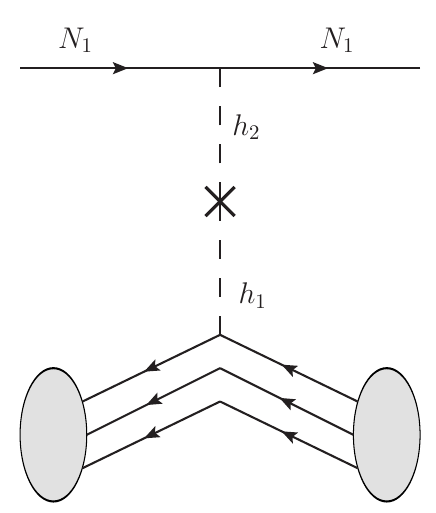}
				\end{center}
				 \caption{DM-nucleon scattering mediated by scalars in scotogenic $U(1)_{B-L}$ model.}
               \label{dd1}
\end{figure}

Apart from the relic abundance constraints from Planck experiment, there exists strict bounds on the dark matter nucleon cross section from direct detection experiments like LUX \cite{Akerib:2016vxi}, PandaX-II \cite{Tan:2016zwf, Cui:2017nnn} and Xenon-1T \cite{Aprile:2017iyp, Aprile:2018dbl}. For right-handed neutrino DM in our model, there are two ways DM can scatter off nuclei: one is mediated by $Z_{B-L}$ gauge boson and the other is mediated by scalars. The scalar mediated interactions occur due to mixing of singlet scalars of the model with the SM Higgs boson. Due to the Majorana nature of DM, the $Z_{B-L}$ mediated diagram contribution to the spin-independent direct detection cross section turns out to be velocity suppressed and hence remains within experimental bounds. The scalar mediated diagram shown in figure \ref{dd1} can however, saturate the latest experimental bounds. For the scalar mediated case, the spin-independent elastic scattering cross-section of DM per nucleon can be written as,
\begin{equation}\label{SI_crosssection}
\sigma_{SI}^{h_{1}h_{2}}=\frac{{\mu_{r}}^{2}}{\pi A^{2}} \left[Z f_{p} + (A-Z)f_{n} \right]^{2}
\end{equation}
where
A and Z are the mass and atomic number of the target nucleus respectively. $\mu_{r}$ is the reduced mass. The interaction strengths of proton $f_{p}$ and neutron $f_{n}$ with DM can be written as,
\begin{equation}
	f_{p,n}=\sum\limits_{q=u,d,s} f_{T_{q}}^{p,n} \alpha_{q}\frac{m_{p,n}}{m_{q}} + \frac{2}{27} f_{TG}^{p,n}\sum\limits_{q=c,t,b}\alpha_{q} 
\frac{m_{p,n}}{m_{q}}\,,
\label{fpn}
\end{equation}
and 
\begin{equation}
 \alpha_{q} = \frac{\lambda_{11}\sin2\gamma}{2\sqrt{2}} \left( \frac{m_{q}}{v}\right) \left[\frac{1}{M_{h_{2}}^{2}}-\frac{1}{M_{h_{1}}^{2}}\right] \,.
  \label{DD4}
 \end{equation}
 In the above Eq.\eqref{fpn}, the $f_{T_{q}}^{p,n}$ are given by $f_{Tu}^{(p)}=0.020\pm 0.004, f_{Td}^{(p)}=0.026\pm0.005, f_{Ts}^{(p)}=0.118\pm0.062, f_{Tu}^{(n)}=0.014\pm0.003, f_{Td}^{(n)}=0.036\pm0.008, f_{Ts}^{(n)}=0.118\pm0.062$~\cite{Ellis:2000ds}. 
 
 Using these, the spin-independent cross section Eq.\eqref{SI_crosssection} can be re-expressed as:
\begin{eqnarray}
\sigma_{SI}^{h_{1}h_{2}} &=& \frac{{\mu_{r}}^{2}}{\pi A^{2}} \left(\frac{\lambda_{11} \sin 2 \gamma}{2\sqrt{2}} \right)^{2}  \left[ \frac{1}{M_{h_{2}^{2}}}-\frac{1}{M_{h_{1}^{2}}} \right]^{2} \nonumber \\
& \times & \left[ Z \left(\frac{m_{p}}{v}\right) \left(f_{Tu}^{p}+f_{Td}^{p}+f_{Ts}^{p}+\frac{2}{9}f_{TG}^{p} \right) + (A-Z) \left(\frac{m_{n}}{v}\right) \left(f_{Tu}^{n}+f_{Td}^{n}+f_{Ts}^{n}+\frac{2}{9}f_{TG}^{n}\right) \right]^{2}. \nonumber\\
\label{SI_cross}
\end{eqnarray}

%%%%%%%%%%%%%%%%%%%%%%%%%%%%%%%%%%%%%%%%%%%%%%%%%%%%%%%%%%%%%%%%%%%%%%%%%%%%%%%%%%%%%%%%%%%%%%%%%%% 
\begin{figure}
\begin{center}
\includegraphics[width = 100mm]{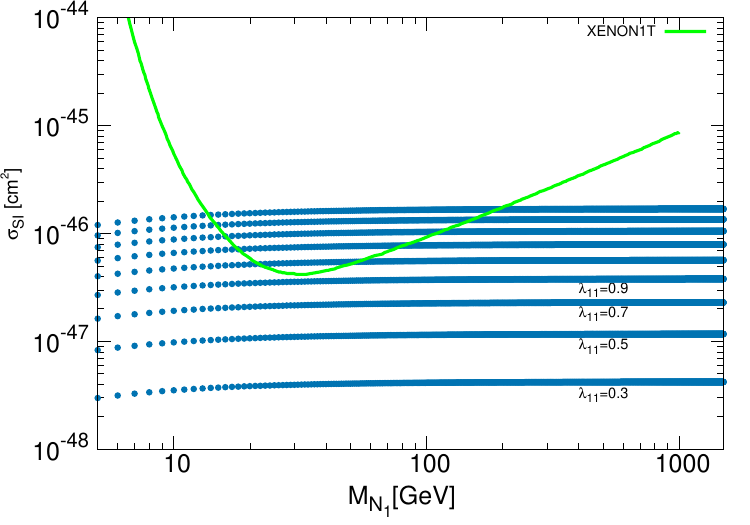}
\end{center}
				 \caption{Spin-independent DM-nucleon scattering cross section mediated by scalars in comparison to the latest Xenon-1T bounds.}
               \label{dd2}
\end{figure}

We show the DM-nucleon cross section mediated by scalars in figure \ref{dd2} in comparison to the latest Xenon-1T bound \cite{Aprile:2018dbl}. 
The only unknown parameter in Eq.\eqref{SI_cross} is $\lambda_{11}$ and $\sin 2 \gamma$. $\sin \gamma$ is taken as 0.1. In figure~\ref{dd2}, the blue points show the spin-independent DM-nucleon cross-section for the values of $\lambda_{11}$ in between $(0.2-2)$ from bottom to top at a step of $0.1$.
As can be seen from this plot, the model remains sensitive to present direct detection experiments, specially when $\lambda_{11} \gtrsim 0.9$. Note that for $\lambda_{11} \lesssim 0.9$ is compatible with bounds from Xenon-1T as well as relic density.

\begin{table}[h!]
%\centering
\vspace{0.3cm}
\begin{tabular}{||l|l|l||}
\hline
$M_{N_1}$ (GeV) & $M_{\eta^{\pm}}$ , M$_{\eta^{0R}}$, M$_{\eta^{0I}}$ (GeV)& $\sigma_{p\,p\,\rightarrow\eta^+\eta^-}$ (pb)\\
\hline 
\hline
100 & 105, 120, 120 & 0.189 \\
\hline
200 & 205, 220, 220 & 1.65 $\times 10^{-2}$  \\
\hline
300 & 305, 320, 320 & 3.46 $\times 10^{-3}$\\
\hline
400 & 405, 420, 420 & 1.04 $\times 10^{-3}$  \\
\hline
500 & 505, 520, 520 & 3.817 $\times 10^{-4}$ \\
\hline
600 & 605, 620, 620 & 1.593 $\times 10^{-4}$ \\
\hline
700 & 705, 720, 720 & 7.286$\times 10^{-5}$  \\
\hline
800 & 805, 820, 820 & 3.568$\times 10^{-5}$ \\
\hline
900 & 905, 920, 920 & 1.828$\times 10^{-5}$\\
\hline
1000 & 1005, 1020, 1020 & 9.794$\times 10^{-6}$ \\
\hline
\end{tabular}
\caption{Production cross sections of
$\eta^{+}\eta^{-}$
from $p\,p$ collisions 
at $\sqrt{s}=14$ TeV LHC. Here we have kept fixed the mass splittings as M$_{\eta^{\pm}} - M_{N_{1}}$=5 GeV and M$_{\eta^{0R}} - M_{\eta^{\pm}}$=M$_{\eta^{0I}} - M_{\eta^{\pm}}$=15 GeV}
\label{Tab:LHC1}
\end{table}

\begin{table}[h!]
%\centering
\vspace{0.3cm}
\begin{tabular}{||l|l|l||}
\hline
$M_{N_1}$ (GeV) & $M_{\eta^{\pm}}$ , M$_{\eta^{0R}}$, M$_{\eta^{0I}}$ (GeV)& $\sigma_{p\,p\,\rightarrow\eta^+\eta^-}$ (pb)\\
\hline 
\hline
100 & 101, 120, 120 & 0.2176 \\
\hline
200 & 201, 220, 220 & 1.782 $\times 10^{-2}$  \\
\hline
300 & 301, 320, 320 & 3.65 $\times 10^{-3}$\\
\hline
400 & 401, 420, 420 & 1.087 $\times 10^{-3}$  \\
\hline
500 & 501, 520, 520 & 3.957 $\times 10^{-4}$ \\
\hline
600 & 601, 620, 620 & 1.647 $\times 10^{-4}$ \\
\hline
700 & 701, 720, 720 & 7.523$\times 10^{-5}$  \\
\hline
800 & 801, 820, 820 & 3.656$\times 10^{-5}$ \\
\hline
900 & 901, 920, 920 & 1.879$\times 10^{-5}$\\
\hline
1000 & 1001, 1020, 1020 & 1.004$\times 10^{-5}$ \\
\hline
\end{tabular}
\caption{Production cross sections of
$\eta^{+}\eta^{-}$
from $p\,p$ collisions 
at $\sqrt{s}=14$ TeV LHC. Here we have kept fixed the mass splittings as M$_{\eta^{\pm}} - M_{N_{1}}$=1 GeV and M$_{\eta^{0R}} - M_{\eta^{\pm}}$=M$_{\eta^{0I}} - M_{\eta^{\pm}}$=19 GeV}
\label{Tab:LHC2}
\end{table}

\begin{table}[h!]
%\centering
\vspace{0.3cm}
\begin{tabular}{||l|l|l||}
\hline
$M_{N_1}$ (GeV) & $M_{\eta^{\pm}}$ , M$_{\eta^{0R}}$, M$_{\eta^{0I}}$ (GeV)& $\sigma_{p\,p\,\rightarrow \eta^{\pm}\eta^{0}}$ (pb)\\
\hline 
\hline
100 & 101.2, 101, 101.2 & 0.2473 \\
\hline
200 & 201.2, 201, 201.2 & 2.057 $\times 10^{-2}$  \\
\hline
300 & 301.2, 301, 301.2 & 4.359 $\times 10^{-3}$\\
\hline
400 & 401.2, 401, 401.2 & 1.341 $\times 10^{-3}$  \\
\hline
500 & 501.2, 501, 501.2 & 5.001 $\times 10^{-4}$ \\
\hline
600 & 601.2, 601, 601.2 & 2.141 $\times 10^{-4}$ \\
\hline
700 & 701.2, 701, 701.2 & 9.938$\times 10^{-5}$  \\
\hline
800 & 801.2, 801, 801.2 & 4.91$\times 10^{-5}$ \\
\hline
900 & 901.2, 901, 901.2 & 2.546$\times 10^{-5}$\\
\hline
1000 & 1001.2, 1001, 1001.2 & 1.367$\times 10^{-5}$ \\
\hline
\end{tabular}
\caption{Production cross sections of
$\eta^{\pm}\eta^{0}$
from $p\,p$ collisions 
at $\sqrt{s}=14$ TeV LHC. Here we have kept fixed the mass splittings as M$_{\eta^{0R}} - M_{N_{1}}$=1 GeV and M$_{\eta^{\pm}} - M_{\eta^{0R}}$=M$_{\eta^{0I}} - M_{\eta^{0R}}$=200 MeV}
\label{Tab:LHC3}
\end{table}

\section{Collider Signatures}\label{collider}
Collider signatures of $U(1)_{B-L}$ models have been discussed extensively in the literature. Since all the SM fermions are charged under this gauge symmetry, the production of $Z_{B-L}$ gauge boson in proton proton collisions can be significant \cite{Okada:2016gsh, Okada:2018ktp, Basso:2008iv}, if the corresponding gauge coupling $g_{B-L}$ is of the same strength as electroweak gauge couplings. Such heavy gauge boson, if produced at colliders, can manifest itself as a narrow resonance through its decay into dileptons, say. The latest measurement by the ATLAS experiment at 13 TeV LHC constrains such gauge boson mass to be heavier than $3.6-4.0$ TeV depending on whether the final state leptons are of muon or electron type \cite{Aaboud:2017buh}. The corresponding bound for tau lepton final states measured by the CMS experiment at 13 TeV LHC is slightly weaker, with the lower bound on $Z_{B-L}$ mass being 2.1 TeV \cite{Khachatryan:2016qkc}. In deriving the bounds for $e^+ e^-, \mu^+ \mu^-$ final states, the corresponding gauge coupling was chosen to be $g_{B-L} \approx 0.28$. Therefore, such bounds can get weaker if we consider slightly smaller values of gauge couplings. For a recent discussion on such signatures, please refer to \cite{Nanda:2017bmi}. For other possible signatures say, right-handed neutrinos in $U(1)_{B-L}$ or similar $Z^{\prime}$ model among others, please see references \cite{Basso:2008iv, Basso:2010pe, Basso:2010yz, Basso:2012sz, Basso:2012ux, Accomando:2013sfa, Okada:2016gsh, Das:2017deo, Das:2017flq, Bandyopadhyay:2018cwu, Deppisch:2018eth, Majee:2010ar}.

Instead of such conventional searches, here we consider two interesting signatures our present version of $U(1)_{B-L}$ model can have. This is related to the production and subsequent decay of the charged component of $Z_2$ odd scalar doublet $\eta$ which can be the NLSP or next to NLSP, while the lightest right handed neutrino is the LSP (DM). The production cross section of charged pairs $\eta^+ \eta^-$ as well as $\eta^{\pm}\eta^{0}$ at 14 TeV proton proton collisions are shown in table \ref{Tab:LHC1}, \ref{Tab:LHC2}, \ref{Tab:LHC3} for different benchmark values of parameters. For this calculation, we implemented the model in {\tt FeynRule}~\cite{Alloul:2013bka} and used {\tt MADGRAPH}~\cite{Alwall:2014hca} for the cross section calculations. Once these particles (i.e., $\eta^\pm$) are produced, they 
live for a longer period before decay to final state particles including DM (i.e., $N_1$) due to phase space suppression. See for instance~\cite{Bhattacharya:2017sml,Bhattacharya:2018fus}. 

A particle like NLSP with sufficiently long lifetime, so that its decay length is of the order of 1 mm or longer, if produced at the colliders, can leave a displaced vertex signature. This vertex, created by the decay of the long lived particle, is located away from the collision point where the decaying particle was created. The final state like charged leptons or jets from such displaced vertex can then be reconstructed by dedicated analysis, some of which in the context of the LHC may be found in \cite{Aaboud:2016dgf, Khachatryan:2016sfv, Aaboud:2017iio}. Similar analysis in the context of upcoming experiment may be found in \cite{Curtin:2017izq, Curtin:2017bxr} and references therein.

Since such signatures are very much clean, one can search for such particles at colliders with relatively fewer events. Here we make some crude estimates at the cross section level and decay length without going into the details of event level analysis. For recent searches of displaced vertex type signatures at the LHC, one may refer to \cite{Aaboud:2017iio, Aaboud:2017mpt}. For a recent discussion on such signatures in type I seesaw model and active-sterile neutrino mixing case, please see \cite{Jana:2018rdf}  and \cite{Cottin:2018nms}.
\begin{figure}[t!]
%\centering
\includegraphics[scale=0.75]{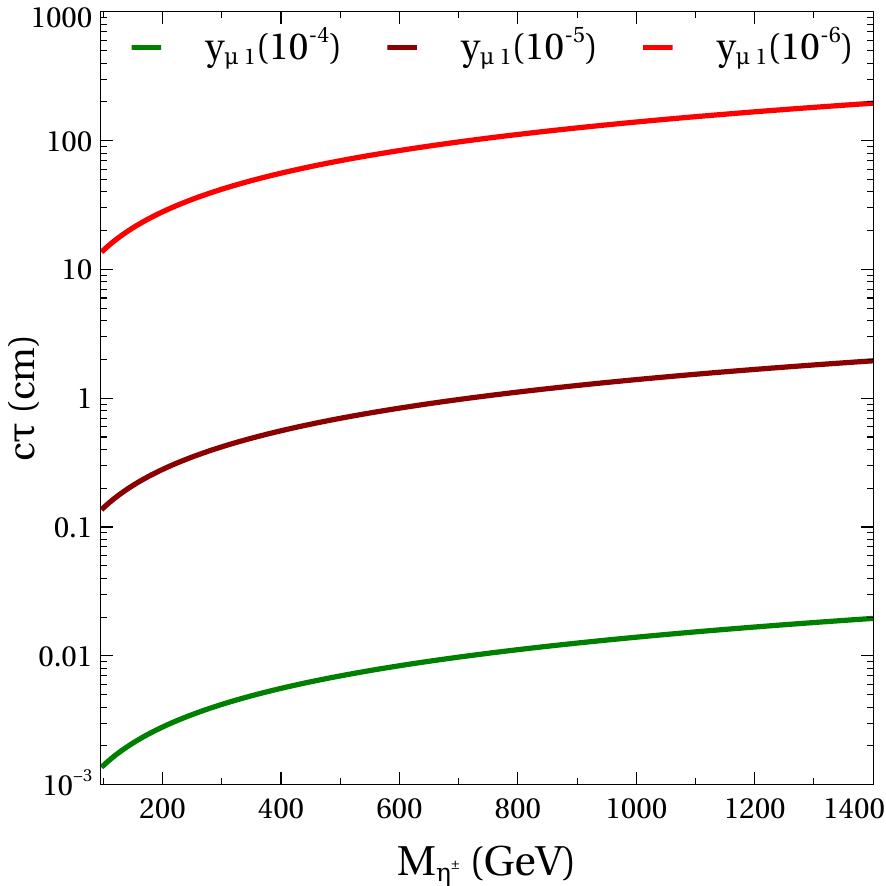}
\caption{Decay length of $\eta^{\pm}\rightarrow N_1 \,\mu$ as a function of $\eta^{\pm}$ mass.}
\label{fig:decay1}
\end{figure}

The decay width of $\eta^\pm$ can be written as 
\begin{equation}
\Gamma_{\eta^{\pm}\rightarrow N_{1} \mu}=\frac{y_{\mu_1}^2\Big(m_{\eta^{\pm}}^2-(m_{N_1}+m_{\mu})^2\Big)}{8 m_{\eta^\pm}\pi}\sqrt{1-\bigg(\frac{m_{N_1}-m_\mu}{m_\eta^\pm}\bigg)^2}\sqrt{1-\bigg(\frac{m_{N_1}+m_\mu}{m_\eta^\pm}\bigg)^2}
\end{equation}
where $y_{\mu 1}$ is the Yukawa coupling of the interaction $\eta^{\pm} N_{1} \mu$. The corresponding decay length as a function of $\eta^\pm$ mass for different benchmark values of $y_{\mu 1}$ are shown in figure \ref{fig:decay1}. At high luminosity LHC, decay length of a few cm can be searched for, if the decaying particle has production cross section of the order a few fb or more \cite{Jana:2018rdf}, which is clearly satisfied for several benchmark masses as shown in table \ref{Tab:LHC1}, \ref{Tab:LHC2}, \ref{Tab:LHC3}. Although such tiny Yukawa couplings required for displaced vertex signatures will not induce any co-annihilations between $N_1$ and the components of $\eta$, we can still have strong co-annihilations due to tau lepton couplings while $\eta^{\pm}$ decay into DM and tau lepton can be kinematically forbidden. In such a case, DM $(N_1)$ can be sufficiently light due to strong co-annihilations via tau lepton sector couplings but at the same time we can have displaced vertex signatures of $\eta^{\pm}$ into first two generation charged leptons. Future proposed experiments like the Large Hadron electron Collider (LHeC), Future Circular electron-hadron Collider (FCC-eh) will be able to search for even shorter decay lengths and cross sections, than the ones discussed here.

\begin{figure}[t!]
%\centering
\includegraphics[scale=0.5]{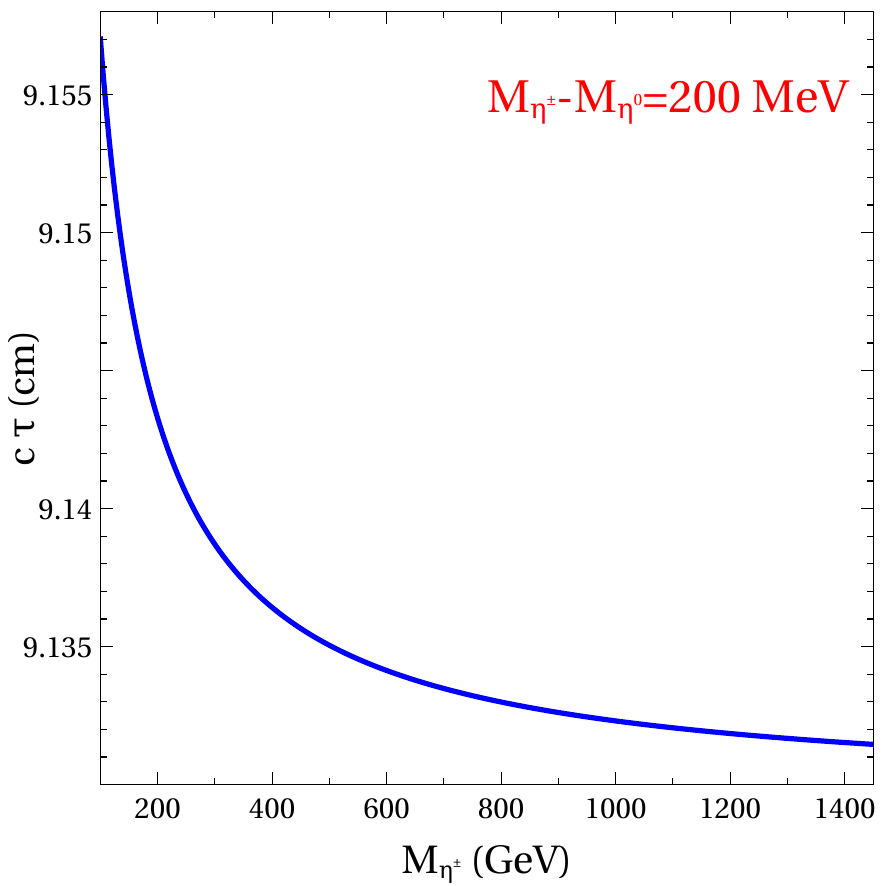}
\includegraphics[scale=0.5]{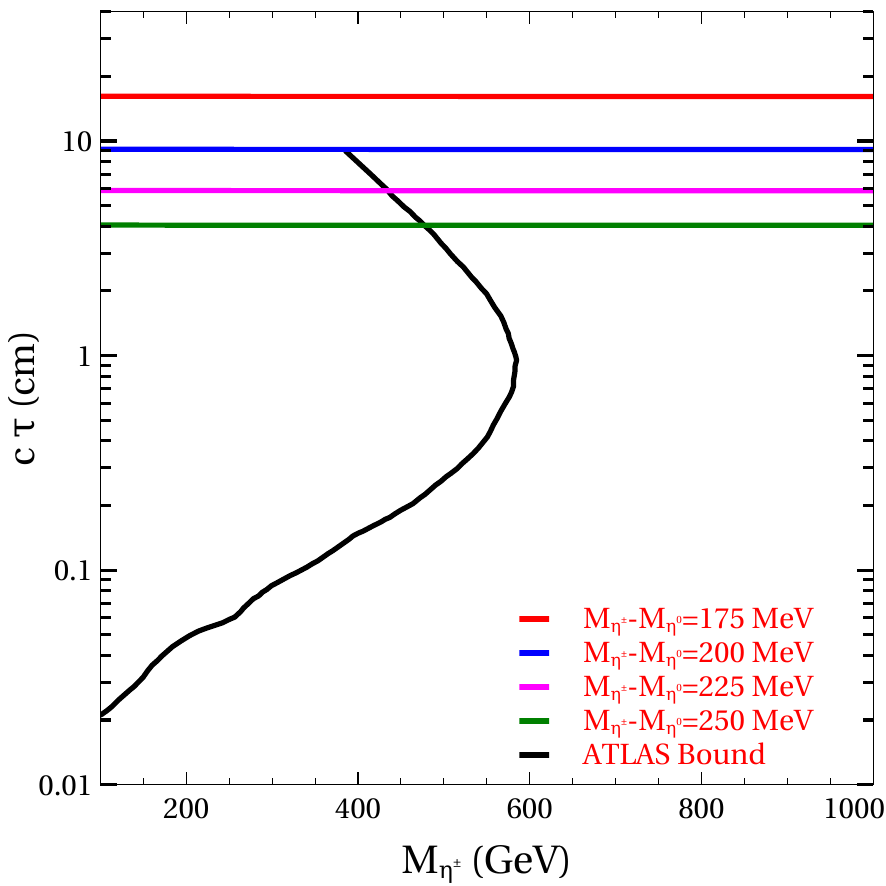}
\caption{Decay length corresponding to the pionic decay $\eta^{\pm}\rightarrow \eta^0\,\pi^\pm$ for fixed mass splitting of 200 MeV (left panel) and its comparison with the ATLAS bound for different benchmark values of mass splitting (right panel).}
\label{fig:decay2}
\end{figure}

\begin{figure}[t!]
%\centering
\includegraphics[scale=0.75]{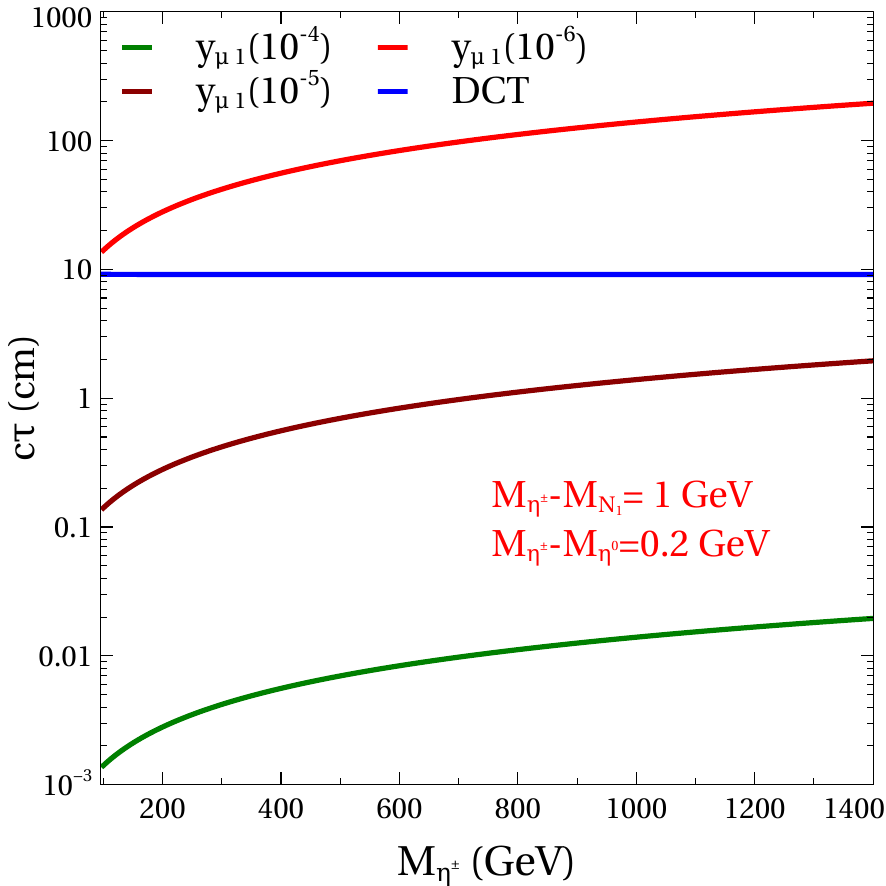}
\caption{Decay length corresponding to the pionic decay $\eta^{\pm}\rightarrow \eta^0\,\pi^\pm$ leading to DCT and its comparison with the decay $\eta^{\pm}\rightarrow N_1 \,\mu$ responsible for displaced vertex signature.}
\label{fig:decay3}
\end{figure}

Another interesting possibility arises when the mass splitting between $\eta^{\pm}$ and $\eta^0$ is very small, of the order of 100 MeV. For such mass splitting, the dominant decay mode of $\eta^{\pm}$ can be $\eta^{\pm} \rightarrow \eta^0 \pi^{\pm}$, if the corresponding Yukawa coupling of $\eta^{\pm} N_{1} l$ vertex is kept sufficiently small for the leptonic decay mode to be subdominant. The corresponding decay width is given by 
\begin{equation}
\Gamma_{\eta^{\pm}\rightarrow \eta^0 \pi^\pm}=\frac{f_{\pi}^2 g^4 }{m_{W}^4}\frac{\Big(m_{\eta^\pm}^2-m_{\eta^0}^2\Big)^2}{512 m_{\eta^\pm}\pi}\sqrt{1-\bigg(\frac{m_{\eta^0}-m_\pi}{m_\eta^\pm}\bigg)^2}\sqrt{1-\bigg(\frac{m_{\eta^0}+m_\pi}{m_\eta^\pm}\bigg)^2}
\end{equation}
where $f_{\pi}$, g, m$_W$ are the form factor, gauge coupling, and W boson mass respectively. Such tiny decay width keeps the lifetime of $\eta^{\pm}_1$ considerably long enough that it can reach the detector before decaying. In fact, the ATLAS experiment at the LHC has already searched for such long-lived charged particles with lifetime ranging from 10 ps to 10 ns, with maximum sensitivity around 1 ns \cite{Aaboud:2017mpt}. In the decay $\eta^{\pm}\rightarrow \eta^0\,\pi^\pm$, the final state pion typically has very low momentum and it is not reconstructed in the detector. On the other hand, the neutral scalar in the final state $\eta^0$ eventually decays into DM and a light neutrino and hence remain invisible throughout. Therefore, it gives rise to a signature where a charged particle leaves a track in the inner parts of the detector and then disappears leaving no tracks in the portions of the detector at higher radii. The corresponding decay length as a function of $\eta^{\pm}$ mass is shown in the left panel plot of figure \ref{fig:decay2}. The right panel plot of figure \ref{fig:decay2} shows a comparison of the decay length in our model with the ATLAS bound \cite{Aaboud:2017mpt}. In figure \ref{fig:decay3}, we show the comparison between the leptonic decay mode and pionic decay mode for different benchmark values of Yukawa couplings.

\section{Conclusions}
\label{sec6}
We have studied a simple extension of the minimal gauged $U(1)_{B-L}$ with three right-handed neutrinos in order to realise fermion singlet dark matter. The minimal model is extended by a scalar doublet $\eta$ and an additional $Z_2$ symmetry so that the right-handed neutrinos and $\eta$ are odd under this $Z_2$ symmetry while all other fields are even. Neutrinos remain massless at tree level but acquires a radiative contribution with the $Z_2$ odd fields going in the loop, in a way similar to scotogenic scenarios. The lightest $Z_2$ odd particle, considered to be the lightest right-handed neutrino, is the dark matter candidate in the model. Due to lepton portal interactions and hence several co-annihilation channels, there exists enlarged parameter space in terms of dark matter mass so that the correct relic abundance is obtained. This is in sharp contrast with minimal fermion singlet dark matter scenarios where relic is usually satisfied only in the vicinity of resonance regions. We also find that the co-annihilation between right-handed neutrino DM and the $Z_2$ odd scalar doublet remains dominant over that between DM and heavier right-handed neutrinos.

Here we note that the DM relic is generated by virtue of both gauge, scalar portal as well as Yukawa interactions of the lightest right handed neutrino. On the contrary, in pure scotogenic model, the fermion DM relic will solely depend upon Yukawa couplings and hence require large values of the latter to enhance the annihilations. As pointed out by the authors of \cite{Lindner:2016kqk}, such large values of Yukawa couplings often destabilise the $Z_2$ symmetric vacuum at a scale below that of the heaviest right handed neutrino thereby making it inconsistent. However, as we can see from the required values of Yukawa couplings shown in figure \ref{deltm_yuk}, \ref{mn1_yuk} in order to satisfy all requirements, we do not have large Yukawa couplings beyond unity, keeping the $Z_2$ symmetric vacuum stable at low energy scale relevant to the desired phenomenology.

After showing the parameter space allowed from relic abundance criteria, we incorporate the constraints from neutrino mass and dark matter direct detection. While the direct detection scattering mediated by the $U(1)_{B-L}$ gauge bosons remain velocity suppressed, the scalar mediated contribution can saturate the current limits on spin-independent direct detection cross section. Since the Yukawa interactions responsible for enhanced co-annihilation of DM with scalar doublet also appear in one loop neutrino mass formula and can lead to charged lepton flavour violation like $\mu \rightarrow e \gamma$ at one-loop, we can tightly constrain them from existing constraints, in addition to the relic bounds. Motivated from collider signature point of view, we consider small mass splitting (less than tau lepton mass ) between DM and scalar doublet (NLSP). Moreover, the tri-linear couplings of the scalar doublet with the first two generations of leptons: $\eta^\pm N_1 e^\mp$ 
and $\eta^\pm N_1 \mu^\mp$ are assumed to be small so that $\eta^\pm$ (the NLSP) after getting produced significantly at the LHC due to electroweak gauge interactions, can give rise to displaced vertex signatures via decaying into muon or electrons. One can also have a disappearing charged track signature where the charged component of the scalar doublet can decay into the neutral component and a pion with too low kinetic energy to get detected. Both these types of signatures are being searched for the LHC and could be a promising way of discovering BSM physics apart from the usual collider prospects of $U(1)_{B-L}$ models. We constrain the parameter space from the requirements of DM relic density, direct detection, light neutrino masses and mixing, MEG 2016 bound on $\mu \rightarrow e \gamma$ and finally from the requirement of producing displaced vertex signatures at the LHC. We find that the model can have discovery prospects at direct search and LFV experiments as well, apart from the LHC signatures. 

\acknowledgments
We thank the organisers of WHEPP XV at IISER Bhopal (14-23 December, 2017), where this work was initiated. DB acknowledges the support from IIT Guwahati start-up grant (reference number:  xPHYSUGI-ITG01152xxDB001) and Associateship Programme of IUCAA, Pune. DN would like to thank Shibananda Sahoo for useful discussions.

%\bibliographystyle{apsrev}
%\bibliography{ref.bib}

\end{document}